\documentclass[showpacs,superscriptaddress,preprintnumbers]{revtex4}
\usepackage{amsfonts}
\usepackage{amssymb}
\usepackage{amsmath}
\usepackage{epsfig}
\usepackage{tabularx}
\usepackage{array}
\usepackage{subfig}
\usepackage{color}
\usepackage{float}
\allowdisplaybreaks
\begin{document}
\title{Dynamical Structure of Einasto Dark Matter Wormholes: Thin-Shell Stability and Particle Transitions}
%\title{Wormholes and nonlinear gravitational dynamics: from geometry to particle motion} 
%\title{Can Einasto Dark Matter Sustains Traversable Wormholes and Particle Dynamics}

\author{M. Yousaf}
\email{myousaf.math@gmail.com}
\affiliation{Department of Mathematics, Virtual University of Pakistan,\\ 54-Lawrence Road, Lahore 54000, Pakistan.} 
\affiliation{Research Center of Astrophysics and Cosmology, Khazar University, Baku, AZ1096, 41 Mehseti Street, Azerbaijan.}

\author{Faisal Javed}
\email{faisaljaved.math@gmail.com (Corresponding~Author)}
\affiliation{College of Transportation, Tongji University, Shanghai 201804, People’s Republic of China}

\author{Gulzoda Rakhimova}
\email{rakhimovagulzoda96@gmail.com}
\affiliation{University of Tashkent for Applied Sciences, Str. Gavhar 1, Tashkent 100149, Uzbekistan}
\affiliation{National Research University TIIAME, Kori Niyoziy 39, Tashkent 100000, Uzbekistan}

\author{Farkhod Botirov}
\email{f.botirov@nuu.uz}
\affiliation{National University of Uzbekistan, Tashkent 100174, Uzbekistan}

\author{Farruh~Atamurotov}
\email{atamurotov@yahoo.com}
\affiliation{Kimyo International University in Tashkent, Shota Rustaveli str. 156, Tashkent 100121, Uzbekistan}


\begin{abstract}
In this manuscript, we develop topologically charged static and spherically symmetric wormhole (WH) configurations in Einstein gravity by employing the Einasto dark matter profile to construct an analytic shape function. The resulting spacetime satisfies the fundamental traversability conditions, while the matter sector exhibits localized violations of energy conditions near the throat, indicating confined exotic matter content. We analyze equilibrium through the TOV framework, revealing a non-equilibrium force competition in which hydrostatic and anisotropic contributions self-organize to sustain static configurations within specific parameter domains. Stability is examined via the anisotropy parameter and further extended to linearized radial perturbations of the surrounding shell, uncovering parameter dependent stability windows that characterize the system's nonlinear response. A detailed dynamical analysis shows that the monopole charge parameter significantly restructures the effective potential landscape, modifying curvature and inducing qualitative transitions in particle motion. The coupled interplay between monopole charge, angular momentum, and particle energy governs a nonlinear transition from tightly wound quasi bound states to unbounded scattering trajectories. while this transition reflects an emergent dynamical phase structure in the WH spacetime. Furthermore, the complexity factor exhibits strong localization near the throat and vanishes asymptotically, indicating that structural complexity is confined to the inner nonlinear regime, whereas a volume integral quantifier is employed to estimate the total exotic matter content required to sustain the configuration.
\end{abstract}

\maketitle

\section{Introduction}

Einstein's formulation of gravitation fundamentally transformed the classical Newtonian picture of gravity from an instantaneous force acting in an absolute background into a geometric manifestation of curved spacetime. The development of special relativity established the observer dependent character of space and time also resolved incompatibility between classical kinematics as well as Maxwellian electrodynamics~\cite{einstein1982created,holton1960origins}. This conceptual scenario subsequently generalized through Einstein's theory of general relativity (EGR), in which distribution of matter and energy determines the curvature of spacetime through the Einstein field equations~\cite{einstein1916foundation}. The geometrical description of gravitation not only accounts for several phenomena inaccessible to Newtonian gravity, but also permits highly nontrivial spacetime configurations such as black holes and wormholes (WHs), while the concept of a WH emerged naturally from investigations of global structure of relativistic spacetimes, whereas terminology itself introduced later by Wheeler~\cite{shinkai2015wormhole}. A WH may be interpreted as a nontrivial spacetime connection joining two asymptotically distinct regions or two widely separated domains of same universe~\cite{morris1988wormholes}. An early realization of a bridge like geometry obtained by Einstein and his collaborator Rosen from maximally extended Schwarzschild solution~\cite{einstein1935particle}. The corresponding Einstein-Rosen bridge, however, does not constitute a physically traversable passage, since its throat is associated with horizon structure and cannot remain open long enough to allow causal communication between two exterior regions. Subsequent investigations demonstrated transient character of such configurations~\cite{wheeler1955geons}, more generally, WH geometries appear as particular solutions of Einstein field equations~\cite{hawking1975large}, as well as their causal, geometrical, and physical properties studied extensively~\cite{visser1995lorentzian}. A major obstacle to constructing a macroscopic traversable WH within EGR is requirement of matter that violates one or more of standard energy conditions, particularly null energy constraints in vicinity of throat, although negative energy configurations may arise through quantum effects, their realization at astrophysical scales remains highly nontrivial~\cite{callan1998brane,visser1995lorentzian}. Nevertheless, the theoretical implications of traversable Lorentzian WHs motivated sustained interest because such geometries may provide unconventional causal connections, distinctive particle trajectories, as well as, under appropriate circumstances, closed timelike curves~\cite{ellis1973ether,bronnikov1973scalar,clement1981einstein,hochberg1998dynamic,simonetti2021sensitive,dai2018new}.

The modern scenario for a humanly traversable WH developed by Morris and his collaborator Thorne~\cite{morris1988wormholes}, who formulated horizon free geometries satisfying explicit throat as well as flare-out requirements while emphasizing necessity of sufficiently controlled tidal forces. The supporting material generally possesses a large radial tension as well as consequently departs from conventional energy conditions, such departures are especially important near throat, where flare-out property is directly related to null energy constraints violation. The causal consequences of traversable WHs also connected with chronology horizons as well as possible generation of closed timelike curves~\cite{kim1991vacuum}, for static configurations, throat can be characterized geometrically as a minimal area surface, whereas dynamical WHs require a more general treatment~\cite{hochberg1998dynamic,hayward1999dynamic}. From an astrophysical perspective, exterior appearance of a WH may share several characteristics with other compact gravitational systems, thereby motivating studies of possible observational signatures as well as particle dynamics around such objects~\cite{hawking1988wormholes,singh2020conformally,mustafa2021traversable,hassan2021traversable,de2023epicyclic,mustafa2023imprints,yousaf2023cylindrical,dai2019observing}. The investigation of WHs is not restricted to a single geometrical or physical environment, while several possible formation mechanisms and configurations explored in compact object as well as extended source scenarios. For instance, competition between gravitational attraction as well as brane tension investigated as a possible mechanism for WH formation, with compact and massive systems providing favorable environments~\cite{dai2020form}. WH configurations possessing cylindrical symmetry also studied as extensions of conventional spherically symmetric scenario; however, construction of cylindrically symmetric solutions with physically satisfactory asymptotic properties remains particularly challenging~\cite{bronnikov2009cylindrical,di2017spin}. These developments illustrate that the matter distribution surrounding a compact geometry is as important as its underlying symmetry, while this observation becomes especially relevant when WHs are considered in galactic environments where dark matter may constitute the dominant gravitating component.

Dark matter is inferred primarily through its gravitational influence on luminous matter, galactic rotation curves, gravitational lensing, as well as dynamics of galaxies and galaxy clusters, and constitutes a substantial fraction of the cosmic matter budget~\cite{salucci2019distribution}. Historically, gravitational evidence for unseen matter can be traced to Zwicky's analysis of galactic clusters~\cite{zwicky1979masses}, subsequent observations of galactic systems provided further evidence for massive nonluminous components; for example, H-I observations of II Zwicky $40$ revealed dynamical masses substantially larger than those inferred from visible matter~\cite{brinks1988dark}. A broad range of observational as well as theoretical investigations has since strengthened the case for dark matter halos around galaxies and compact astrophysical systems~\cite{roberts1978twenty,roszkowski2018wimp,de2021epicyclic}, while this naturally raises question of whether realistic dark matter distributions may themselves provide an environment capable of supporting nontrivial spacetime structures such as traversable WHs. This possibility motivated construction of WH geometries using phenomenological dark matter halo profiles, however Rahaman \emph{et al.} demonstrated that galactic halos described through the Navarro-Frenk-White distribution and rotation curve information may admit traversable WH configurations~\cite{rahaman2014possible}. Related investigations subsequently examined dark matter supported WHs within different gravitational settings~\cite{sharif2016wormhole}, including isothermal and Navarro-Frenk-White galactic halo models~\cite{rahaman2016study}, generalized Navarro-Frenk-White descriptions of dark matter dominated galaxy Dragonfly $44$ studied in~\cite{islam2019formation}, and spherically symmetric WH configurations supported by isotropic halo matter~\cite{xu2020possibility}. These studies indicate that detailed radial distribution of dark matter can significantly influence throat geometry, energy condition behavior, as well as overall viability of a WH spacetime, consequently, employing observationally motivated halo profiles provides a useful bridge between mathematical WH solutions and astrophysical environments. Alongside geometrical viability, internal organization as well as dynamical stability of self-gravitating configurations provide additional criteria for assessing their physical character. Herrera introduced complexity factor for static self-gravitating systems by combining influence of density inhomogeneity as well as pressure anisotropy through orthogonal splitting of Riemann tensor~\cite{herrera2018new}. while the same general program of understanding anisotropic and dissipative gravitating systems also developed through detailed analyses of gravitational collapse, causal transport, bulk as well as shear viscosity, heat flow, and radiative processes~\cite{herrera2009dynamics,herrera2002relativistic,herrera2004dynamics}. The complexity factor approach subsequently extended to a variety of relativistic astrophysical systems and modified gravitational scenarios~\cite{bhatti2021electromagnetic,yousaf2021quasi}. for a WH sustained by anisotropic matter, such quantities are particularly informative because density gradients and unequal radial and tangential stresses are intimately connected with both the exotic matter content and the stability of throat.

Among the available dark matter distributions, the Einasto profile provides a flexible nonsingular representation in which logarithmic slope changes continuously with radial distance which was originally employed in modelling galactic matter distributions, including the mass distribution associated with M31~\cite{einasto1969andromeda}, and subsequently applied to several galactic systems. Numerical studies of cold dark matter halos demonstrated that the inferred inner density structure can be highly sensitive to numerical resolution and treatment of small scale substructure~\cite{moore1998resolving}, while these considerations motivate exploration of smooth alternatives capable of describing realistic galactic density distributions. The successful application of Einasto-type distributions to galaxies such as M31, M32, M87, the Milky Way, and dwarf systems makes this profile particularly suitable for investigating possible connection between dark matter halos and relativistic compact geometries~\cite{einasto1969andromeda,einasto1972galactic}. Motivated by these developments, in present work we construct as well as investigate a class of static, spherically symmetric, topologically charged traversable WHs within Einstein gravity using the Einasto dark matter density profile. In contrast to treating dark matter distribution merely as an auxiliary matter source, we employ the Einasto profile directly in gravitational field equations to derive an analytical WH shape function. The resulting geometry incorporates contribution of a global monopole through the parameter $\xi$, thereby allowing combined influence of dark matter distribution and topological charge on WH geometry to be studied. The throat condition, flare-out requirement, normalized shape function, and related geometrical constraints are examined to establish traversability. The corresponding energy density, radial pressure, and tangential pressure are then analyzed together with the NEC, WEC, SEC, as well as DEC in order to identify regions in which exotic matter is required, in particular, model allows NEC violating contribution to be localized predominantly around throat while conventional matter behavior is progressively recovered away from the central region.

The analysis is further extended beyond construction of background WH solution, while the equilibrium properties of anisotropic configuration are investigated through the Tolman-Oppenheimer-Volkoff equation, pressure anisotropy, and average pressure. We then construct a thin-shell configuration by joining the Einasto based WH interior to a fuzzy black hole exterior and study its response under linearized radial perturbations. This enables identification of stable as well as unstable domains as functions of monopole charge $\xi$ and the Einasto parameters $\gamma_1$, $\beta$, and $\rho_0$, while the intrinsic spatial geometry is visualized through embedding diagrams, whereas timelike geodesics and escape trajectories are employed to investigate the dynamical influence of the monopole charge, particle angular momentum $L$, and energy $E$. In particular, these quantities govern the transition between tightly wound trajectories near the WH and unbounded scattering type motion. We additionally examine the active gravitational mass, the complexity factor $Y_{TF}$, as well as the radial and tangential equation of state parameters, finally, the volume integral quantifier (VIQ) is employed to determine integrated amount of NEC violating matter required to sustain the WH and to assess its dependence on the throat radius and Einasto profile parameters. In this way, the present study combines geometrical construction, matter characterization, equilibrium analysis, thin-shell stability, particle dynamics, structural complexity, and exotic matter quantification within a single Einasto supported topologically charged WH scenario. The manuscript is organized as follows. Section~\ref{BasicFormalism} presents the Einstein gravity framework, the global monopole background, and the basic equations governing the static spherically symmetric WH configuration. In Sec.\ref{ShapeFunction}, the Einasto dark matter density profile is employed to derive the analytical shape function and the fundamental traversability conditions are examined. Section~\ref{PhysicalFeatures_EnergyConstrains} discusses the matter variables as well as the associated energy conditions, whereas Sec.~\ref{Stability} investigates equilibrium through the TOV equation, anisotropy parameter, and average pressure. In Sec.~\ref{TSW}, the thin-shell formalism is developed by matching the WH interior with the fuzzy black hole exterior, and Sec.~\ref{Tinshel_stability} is devoted to the stability of the resulting shell under linearized radial perturbations. Section~~\ref{EmbeddingDiagrams} presents the embedding geometries for different model parameters, while Sec.~\ref{geometry} investigates timelike geodesics and particle escape trajectories. In Sec.~\ref{MassRadius_ComplexityFactor}, we analyze the active gravitational mass, structural complexity through $Y_{TF}$, and the radial and tangential equation of state parameters. Section~\ref{ExoticMatter_VIQ}  quantifies the total amount of exotic matter through the VIQ approach, and the principal results of the study are summarized in last section.

\section{Basic Theoretical Background and Formalism for Wormhole Model}\label{BasicFormalism}

The foundation of Einstein's theory of relativity is provided by the Einstein field equations, which relate the geometry of spacetime to the distribution of matter and energy which form a coupled system of ten nonlinear partial differential equations, involving both hyperbolic and elliptic features. Their mathematical structure is built from geometrical quantities such as the metric tensor, the Ricci tensor, the Ricci scalar, and other curvature related terms, while in compact tonsorial form, the Einstein field equations can be expressed as
\begin{align}\label{b1}
G_{\nu\eta}=\mathcal{K} T_{\nu\eta}~~\text{where}~~\mathcal {K}=\frac{8 \pi G}{c^4}~~\text{and}~~G_{\nu\eta}\equiv R_{\nu\eta}- \frac{1}{2}Rg_{\nu\eta}, 
\end{align}
here, in our work the value of the coupling constant $\mathcal {K}=8\pi$ is considered throughout calculations, while the $G_{\nu\eta}$ is the Einstein's tensor and in the expression above, $G_{\nu\eta}$ represents the curvature of spacetime as determined by the metric tensor $g_{\nu\eta}$, while $T_{\nu\eta}$ denotes the energy and matter distribution within that spacetime. The spacetime geometry generated by a global monopole was first obtained by Barriola and his collaborators~\cite{barriola1989gravitational} and the corresponding line element is given by
\begin{align}\label{1}
d\mathcal{S}^2 = -\left(1 - 8\pi \mathcal{G}\varpi^2 - \frac{2\mathcal{M}}{r}\right) dt^2 
+ \frac{dr^2}{\left(1 - 8\pi \mathcal{G}\varpi^2 - \frac{2\mathcal{M}}{r}\right)} 
+ r^2 \left( d\theta^2 + \sin^2\theta\, d\phi^2 \right),
\end{align}
where $\mathcal{M}$ is a constant which, in the flat spacetime limit, may be interpreted approximately as the mass of the monopole core $\mathcal{M}_{\text{core}}$ and the parameter $\varpi$ denotes the symmetry breaking scale. While $\mathcal{G}$ represents Newton's gravitational constant. If the mass contribution is neglected and suitable rescalings of the temporal and radial coordinates are introduced, the geometry of a point like global monopole can be written in the form~\cite{bezerra2001physics,ahmed2022relativistic}
\begin{align}\label{1additional}
d\mathcal{S}^2 = -dt^2 + \frac{dr^2}{\xi^2} + r^2 \left( d\theta^2 + \sin^2\theta\, d\phi^2 \right),
\end{align}
where
\[
\xi^2 = 1 - 8\pi \mathcal{G}\varpi^2.
\]
Here, $\xi$ may be regarded as an effective monopole parameter, since it is directly related to the symmetry breaking energy scale $\varpi$, while the presence of the global monopole gives rise to a solid angle deficit, $\Delta\Omega = 32\pi^2\mathcal{G}\varpi^2$. Therefore modifies the spacetime geometry by introducing a conical structure. A detailed discussion of the geometrical as well as physical properties of such backgrounds can be found in~\cite{ahmed2022relativistic}. Inspired by these features, we consider a Morris-Thorne-type WH geometry incorporating the topological contribution of a global monopole charge. The matter source is assumed to be a relativistic fluid distribution in a static and spherically symmetric spacetime. In Schwarzschild coordinates $(x^0,x^1,x^2,x^3)=(t,r,\theta,\phi)$, the proposed line element is taken as
\begin{align}\label{2}
d\mathcal{S}^2 = -e^{2A_{1}(r)} dt^2 + \frac{dr^2}{\xi^2\left(1 - \frac{B(r)}{r}\right)} + r^2 \left( d\theta^2 + \sin^2\theta\, d\phi^2 \right),
\end{align}
where $A_1(r)$ represents the redshift function as well as $B(r)$ denotes the WH shape function. The parameter $\xi$ plays the role of a point like monopole charge in topologically charged WH configurations~\cite{ahmed2023topologically,ahmed2024schwarzschild}. Its inclusion modifies the angular sector through a deficit angle approximately given by $\delta\varphi \simeq 2(1-\xi)\pi$. Global monopole geometries as well as their extensions investigated in several gravitational scenarios, which  include monopole fields in topologically charged modified gravity~\cite{soares2023gravitational}, nonlinear $\rho$ models coupled with EiBI theory~\cite{nascimento2020nonlinear}, global monopole solutions in $f(R)$ gravity~\cite{morais2012gravitational}, and monopole charged black hole backgrounds in the context of thermodynamics and gravitational lensing~\cite{man2015analytical}. Similarly, monopole supported traversable WH geometries also explored in $f(Q)$ gravity~\cite{tayde2024exploring}, 4-dimensional topological WH models~\cite{barriola1989gravitational}, and higher-dimensional monopole induced WH scenarios~\cite{ahmed2024five}. For a traversable WH, the redshift function must remain finite throughout the spacetime in order to avoid the formation of horizons. A simple as wll as physically useful choice is to assume a constant redshift function, which implies
\[
A_1'(r)=0.
\]
Therefore, without loss of generality, we set $A_1(r)=0$. Under this assumption, the metric reduces to
\begin{align}\label{3a1}
d\mathcal{S}^2 = -dt^2 + \frac{dr^2}{\xi^2\left(1 - \frac{B(r)}{r}\right)} + r^2 \left( d\theta^2 + \sin^2\theta\, d\phi^2 \right).
\end{align}
This choice considerably simplifies the Einstein field equations and allows a clearer analytical investigation of the WH configuration, in the following sections, we examine the corresponding energy conditions for different well-known choices of the shape function $B(r)$. 
The coordinate ranges are taken as \( r \in [a_{0}, \infty) \), \( -\infty < t < +\infty \), and \( 0 \leq \phi< 2\pi \), where \( a_{0} \) denotes the throat radius.  
Expressing the line element as \( d\mathcal{S}^2 = g_{\nu \eta}\, dx^{\nu } dx^{\eta} \), the metric tensor and its inverse are given by
\[
g_{\nu \eta} = \operatorname{diag}\left(-1, \frac{r}{\xi^2 (r -B(r))}, r^2, r^2\sin^2\theta \right), \quad 
g^{\nu \eta} = \operatorname{diag}\left(-1, \xi^2\left(1 - \frac{B(r)}{r}\right), \frac{1}{r^2}, \frac{1}{r^2\sin^2\theta}\right).
\]
The Ricci scalar corresponding to the metric~\eqref{3a1} is obtained as
\begin{align}\label{3a2}
R = \frac{2 - 2\xi^2 + 2\xi^2 B'(r)}{r^2}.
\end{align}
The energy–momentum tensor (EMT) describing the anisotropic matter content supporting WH is given by
\begin{align}\label{3a3}
T_{\nu \eta} = (\rho + P_{tan})\mathcal{X}_\nu  \mathcal{X}_\eta - (P_{tan} - P_{rad})\mathcal{X}_\nu  \mathcal{X}_\eta + P_{tan} g_{\nu \eta},
\end{align}
where \( \rho \) denotes the energy density, \( \mathcal{X}^\nu  \) is the four-velocity vector, and \( \mathcal{X}^\nu  \) represents the unit radial spacelike vector. These quantities satisfy the standard normalization and orthogonality relations:
\begin{align}
& \mathcal{X}^\nu  = (1,0,0,0),~~\mathcal{X}^\nu  \mathcal{X}_\nu  = 0, \quad \mathcal{X}^\nu  \mathcal{X}_\nu  = -1, \\
& \mathcal{X}^\nu  \mathcal{X}_\nu  = 1,~~\mathcal{X}^\nu  = \left(0, \xi \sqrt{1 - \frac{B(r)}{r}}, 0, 0\right).
\end{align}
The variables related to the matter, namely $\rho$, $P_{rad}$, and $P_{tan} $, can be explicitly expressed by utilizing the Einstein field equation provide in Eqs.\eqref{b1} as,
\begin{align}\label{a4}
\rho &= \frac{-\xi ^2+\xi ^2 B'(r)+1}{r^2},
\\\label{a5}
P_{rad}& = -\frac{\xi ^2 B(r)+\xi ^2 (-r)+r}{r^3},
\\\label{a6}
P_{tan}& = \frac{\xi ^2 \left(B(r)-r B'(r)\right)}{2 r^3}.
\end{align}
It is noticed that in the above equations, there are four unknowns $\rho$,~$P_ {rad}$,~$P_ {tan}$, and $B(r)$. To discover the solution to the WH problem, various methods can be employed due to the surplus of unknowns compared to the number of equations. Below, we present the proposed action plan to address the field equations. It is very interesting to analyze at which spacetime arena, the energy conditions are breached within time-independent, spherically symmetric, $4D$ spacetime configurations. This violation of energy conditions is a notable phenomenon within this specific spacetime framework. The flaring out condition determines whether these conditions are violated \cite{morris1988wormholes}. Furthermore, in higher-dimensional theories, the only places where energy conditions violations can be avoided or satisfied may be near the throat of the WH. 

\section{Determination of the WH Shape Function from the Einasto Profile and Basic Criteria Checks}\label{ShapeFunction}

A thorough understanding of large scale cosmic structures, such as dark matter halos surrounding galaxy clusters, requires accurate modeling of their density distributions, while cosmological N-body simulations suggest that dark matter halos can be effectively represented using three parameter density profiles \cite{merritt2006empirical,hayashi2008understanding}. Among these, the Einasto-density profile emerged as a powerful tool for describing dark matter halos \cite{gao2008redshift,navarro2010diversity,de2019estimation}. The Einasto-density profile also found applications in the study of electromagnetic fields and fuzzy WHs \cite{almatroud2025electromagnetic}, as well as in the construction of WHs using various dark matter density distributions \cite{yousaf2025wormholes}. Furthermore, it provides insights into the central regions of spiral galaxies and the formation of early cosmic structures, including anisotropic WHs analyzed via the Minimal Geometric Deformation technique in the presence of dark matter halos \cite{gadotti2009structural,almatroud2025decoupling}, whereas the three-parameter, non-singular Einasto-density profile offers an accurate and flexible representation of dark matter halos \cite{navarro1997universal}. The literature offers several established methodologies for developing and examining WH models \cite{hawking1988wormholes,singh2020conformally,hassan2021traversable,yousaf2023cylindrical}. The Einasto-density profile is defined through its logarithmic slope, which follows a power law form \cite{retana2012analytical,baes2022einasto}:
$\aleph(r) = \frac{d \ln \rho}{d \ln r} \propto r^{\left\{\frac{1}{\gamma_{1}}\right\}},$ where $\gamma_{1}$ is the Einasto index, a free parameter controlling the curvature of the profile. Integrating this slope leads to the general density function:
$\ln \left( \frac{\rho(r)}{\rho_s} \right) = -a_n \left( \left( \frac{r}{r_s} \right)^{\left\{\frac{1}{\gamma_{1}}\right\}} - 1 \right),$ where $r_s$ denotes the scale radius containing half of the total mass, $a_n$ is a dimensionless parameter ensuring proper normalization, $\rho_s$ is the density at $r_s$, and the central density is $\rho_0 = \rho_s e^{a_n}$. A commonly used representation of the Einasto-density profile for dark matter halos is:
$\ln \left( \frac{\rho(r)}{\rho_{-2}} \right) = -2 \gamma_1 \left( \left( \frac{r}{r_{-2}} \right)^{\left\{\frac{1}{\gamma_{1}}\right\}} - 1 \right),$ where $r_{-2}$ and $\rho_{-2}$ correspond to the radius and density at which the logarithmic slope of the density equals $-2$. Equivalently, the density profile can be written as
\begin{align}\label{a15}
\rho(r) = \rho_0 \, e^{-\left\{\frac{r}{\beta}\right\}^{\left\{\frac{1}{\gamma_{1}}\right\}}},
\end{align}
with the scale length $\beta$ and central density $\rho_0$ defined by
$\beta = \frac{r_s}{a_n^{\gamma_1}} = \frac{r_{-2}}{(2 \gamma_1)^{\gamma_1}}, \quad \rho_0 = \rho_s e^{a_n} = \rho_{-2} e^{2 \gamma_1}.$
For the Einasto-density profile to accurately represent real galactic structures, the density function and associated quantities must satisfy certain physical conditions \cite{einasto1969galactic}: $\rho(r)$ must be strictly positive and finite throughout the spatial domain, while the density should smoothly decrease to zero as $r \to \infty$, avoiding unphysical asymptotic behavior and the total mass, effective radius, and multipole expansions must be finite, whereas all descriptive functions, including mass profiles and gravitational potential, should be continuous without sudden jumps or discontinuities. In the context of traversable WHs, the presence of exotic matter leads to a violation of the null energy constraints. The stability of such WHs can be analyzed either by gradual collapse or by considering configurations with sufficiently large throats, however to derive the corresponding shape function $B(r)$, we combine the Einstein field equation with the Einasto-density profile from Eq.~\eqref{a15}, leading to the differential equation:
\begin{align}\label{a18}
\frac{-\xi^2 + \xi^2 B'(r) + 1}{r^2} = \rho_0 \, e^{-\left\{\frac{r}{\beta}\right\}^{\left\{\frac{1}{\gamma_{1}}\right\}}}.
\end{align}
Solving Eq.~\eqref{a18} yields the WH shape function:
\begin{align}\label{a19}
B(r) = A_{2} + \frac{\gamma_{1} \rho_0 (-r^3) \left\{\frac{r}{\beta}\right\}^{-3\gamma_{1}} \Gamma\left(3\gamma_{1}, \left\{\frac{r}{\beta}\right\}^{\left\{\frac{1}{\gamma_{1}}\right\}}\right) + \xi^2 r - r}{\xi^2},
\end{align}
where $A_{2}$ is the integration constant. Applying the boundary condition at WH throat, $B(a_0) = a_0$, determines $c$ as
\begin{align}\label{a20}
A_{2} = \frac{\gamma_{1} \rho_0 a_0^3 \left\{\frac{a_0}{\beta}\right\}^{-3\gamma_{1}} \Gamma\left(3\gamma_{1}, \left\{\frac{a_0}{\beta}\right\}^{\left\{\frac{1}{\gamma_{1}}\right\}}\right) + a_0}{\xi^2}.
\end{align}
This derivation provides a self consistent expression for the WH shape function $B(r)$ based on the Einasto density profile, ensuring a physically meaningful and mathematically robust WH geometry suitable for further analysis of stability and energy conditions.
The ultimate form of $B(r)$ (i.e. shape function) is provided as,
\begin{align}\nonumber&
B(r)=\frac{1}{\xi ^2}\bigg\{\rho_0 \left(\gamma_{1} a_{0}^3 \left(\left(\frac{a_{0}}{\beta}\right){}^{1/\gamma_{1}}\right){}^{-3 \gamma_{1}} \Gamma \left(3 \gamma_{1},\left(\frac{a_{0}}{\beta}\right){}^{1/\gamma_{1}}\right)-\gamma_{1} r^3 \left(\left(\frac{r}{\beta}\right)^{1/\gamma_{1}}\right)^{-3 \gamma_{1}} \Gamma \left(3 \gamma_{1},\left(\frac{r}{\beta}\right)^{1/\gamma_{1}}\right)\right)
\\\label{a21}&+\left(\xi ^2-1\right) r+a_{0}\bigg\}.
\end{align}

\begin{figure*}[htbp]
\centering
\subfloat[]{{\includegraphics[height=2.8 in, width=3.4 in]{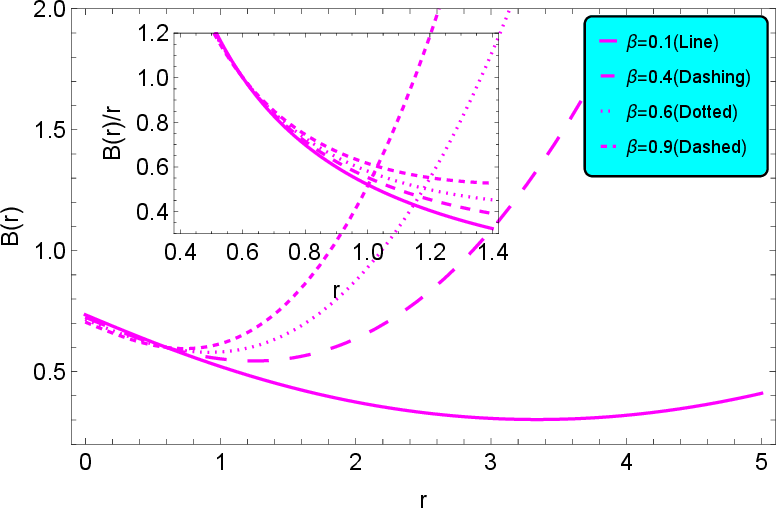}}}
\qquad
\subfloat[]{{\includegraphics[height=2.8 in, width=3.4 in]{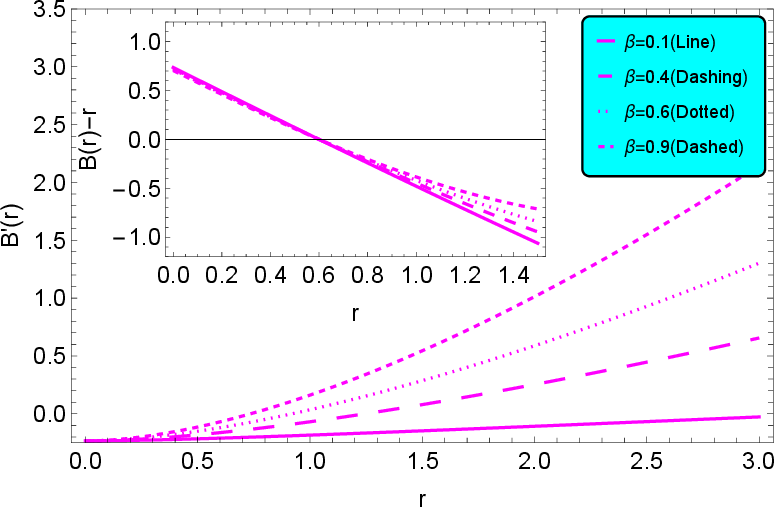}}}
\qquad
\subfloat[]{{\includegraphics[height=2.8 in, width=3.4 in]{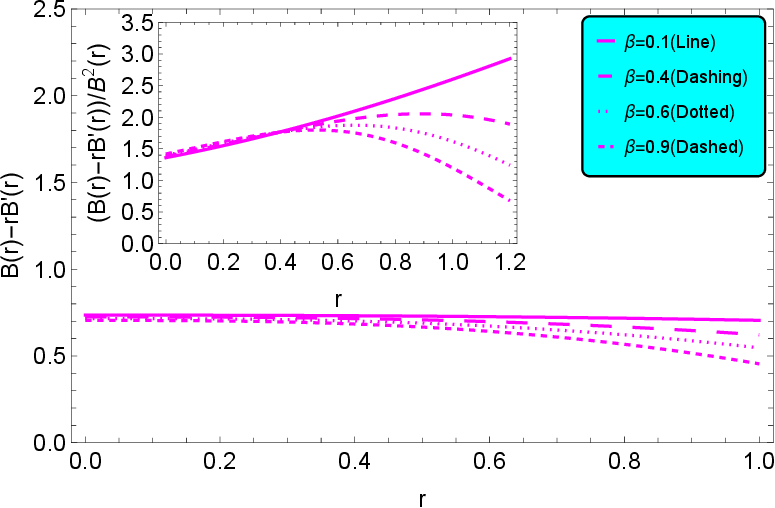}}}
\caption{Plots of shape function \eqref{a21} at 0.6, variation of the shape function itself plotted in main panel (a), inner small panel (a) illustrates the normalized ratio of \eqref{a21} to radial coordinate. Main panel (b) presents the derivative of \eqref{a21} indicating its rate of change, and difference between the shape function and the radial coordinate highlighting the deviation from a linear profile is in inner small panel (b). While difference between the shape function and the product of the radial coordinate with its derivative capturing the interplay of linear and nonlinear effects in main panel (c) and inner small panel (c) shows the normalized flare-out condition.} 
\label{flareout1}
\end{figure*}
The behavior of \eqref{a21} and its instantaneous rate of change with respect to \(r\) is depicted in Figs.~\ref{flareout1}, while all main and inner  panels provide a detailed graphical representation of the constructed shape function using the Einasto dark matter pensity profile as defined in Eq.~\eqref{a15}. Main panel (a) of Figs.~\ref{flareout1} shows the variation of \(B(r)\) vs \(r\), indicating that the function remains positive and increases smoothly. This variation ensuring the proper geometrical structure of constructed WH model and inner small panel (a) of Figs.~\ref{flareout1} presents the normalized ratio \(\frac{B(r)}{r}\). This normalized ratio characterizes the asymptotic behavior of the shape function and ratio decreases with increasing \(r\), satisfying the fundamental WH condition \(\frac{B(r)}{r} < 1\) for \(r > a_0\). The first derivative \(B'(r)\), illustrated in main panel (b) of Fig.~\ref{flareout1}, provides insight into the rate of change of the shape function and plays a crucial role in verifying the flare-out condition at the throat for specific values of the Einasto index $\gamma_{1}$ and different global monopole parameter settings. Main panels (c) and its inner small panel further explore different combinations and normalized forms of the shape function to highlight specific geometric characteristics of the WH. Inner small panel (b) depicts the difference \(B(r)-r\), emphasizing the deviation of the shape function from the radial coordinate, which becomes negative beyond the throat, indicating the open geometry of the WH. Main panel (c) illustrates \(B(r)-rB'(r)\), capturing the interplay between the linear and nonlinear contributions of the Einasto-density profile model. Finally, inner small panel (c) of Fig.~\ref{flareout1} presents the normalized flare-out expression \(\frac{B(r)-rB'(r)}{B(r)^2}\), which is essential for confirming the satisfaction of the flare-out condition necessary for a traversable WH. Consequently, these graphical analyses demonstrate that the constructed Einasto-density profile based shape function satisfies all the required geometric constraints, ensuring a smooth, physically consistent, and traversable WH configuration for a specific range of the Einasto index $\gamma_{1}$ near the WH throat.
The WH metric takes the following form for the previously derived shape function.
\begin{align}\nonumber
d\mathcal{S}^2&=g_{\nu \beta} dx^\nu dx^\beta=-dt^2+\xi^{-2}\bigg[1-\frac{1}{r\xi ^2}\bigg\{\rho_0 \bigg(\gamma_{1} a_{0}^3 \bigg(\bigg(\frac{a_{0}}{\beta}\bigg){}^{1/\gamma_{1}}\bigg){}^{-3 \gamma_{1}} \Gamma \bigg(3 \gamma_{1},\bigg(\frac{a_{0}}{\beta}\bigg){}^{1/\gamma_{1}}\bigg)+a_{0}
\\\label{a22}
&-\gamma_{1} r^3 \bigg(\bigg(\frac{r}{\beta}\bigg)^{1/\gamma_{1}}\bigg)^{-3 \gamma_{1}} \Gamma \bigg(3 \gamma_{1},\bigg(\frac{r}{\beta}\bigg)^{1/\gamma_{1}}\bigg)\bigg)+\bigg(\xi ^2-1\bigg) r\bigg\}\bigg]^{-1}dr^{2}+r^2\left( d\theta^2 + \sin^2\theta\, d\phi^2 \right) ,
\end{align} 

\section{Some Physical Features and Energy Constraints of Fuzzy Wormholes}\label{PhysicalFeatures_EnergyConstrains}

\begin{figure}[H]
\centering
\subfloat[]{{\includegraphics[height=2.8 in, width=3.4 in]{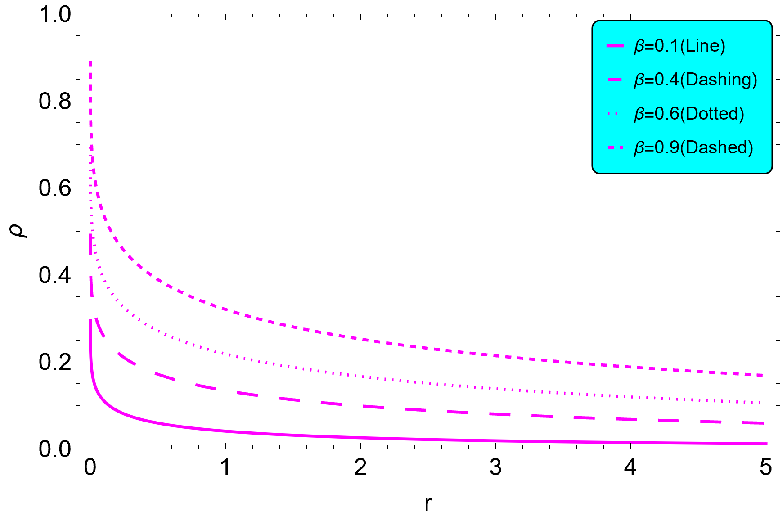}}}
\qquad
\subfloat[]{{\includegraphics[height=2.8 in, width=3.4 in]{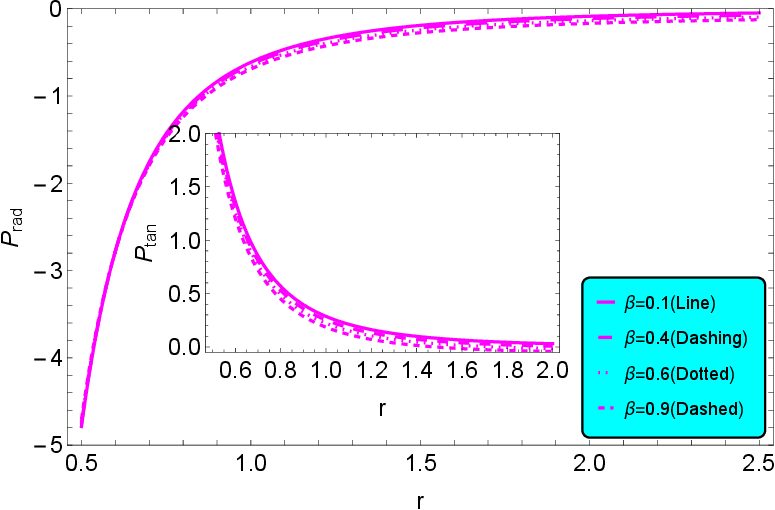}}}
\caption{Behavior of $\rho(r)$ (panel (a)), $P_{rad}(r)$, and $P_{tan}(r)$ (panel (b)) at $a_0 = 0.6$, $\rho_0 $=0.3(Line), $\rho_0 $=0.5(Dashing), $\rho_0 $=0.7(Dotted), $\rho_0 $=0.9(Dashed), and $\gamma_{1}=3.33$.}
\label{energydensity}
\end{figure}

\begin{figure}[H]
\centering
\subfloat[]{{\includegraphics[height=2.8 in, width=3.4 in]{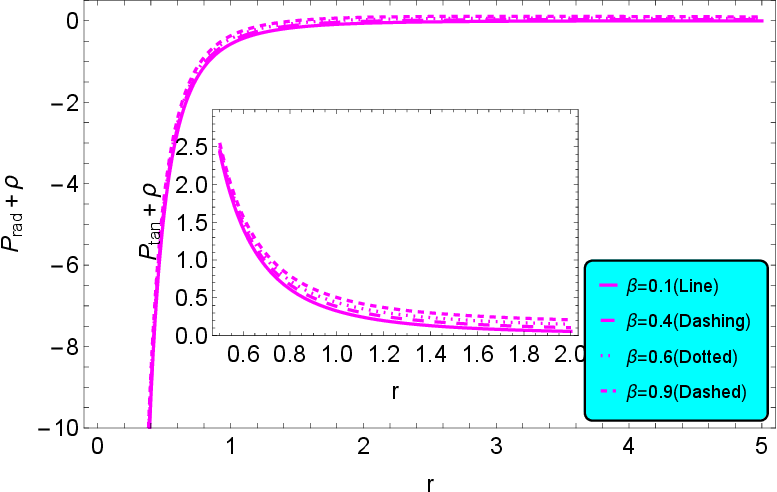}}}
\qquad
\subfloat[]{{\includegraphics[height=2.8 in, width=3.4 in]{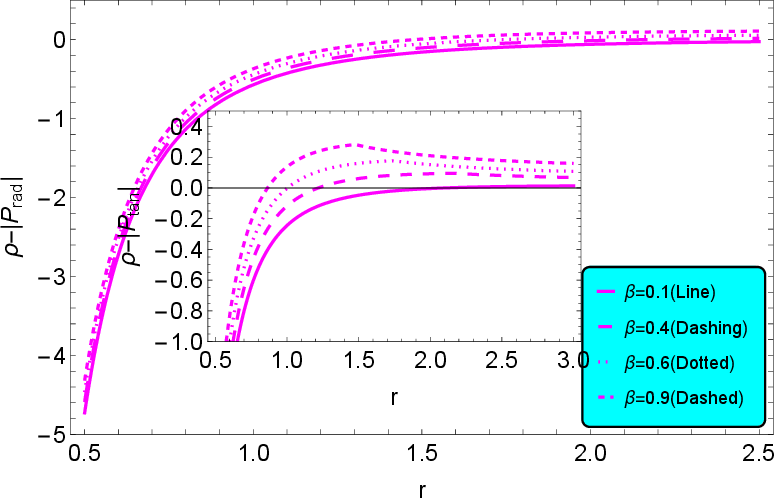}}}
\caption{Behavior of NECs (panel (a)) and DECs (panel (b)) at $a_0 = 0.6$, $\rho_0 $=0.3(Line), $\rho_0 $=0.5(Dashing), $\rho_0 $=0.7(Dotted), $\rho_0 $=0.9(Dashed), and $\gamma_{1}=3.33$.}
\label{NEC}
\end{figure}

Energy conditions play a fundamental role in determining the physical feasibility and matter distribution of WH geometries, whereas they provide mathematical constraints on how energy density and pressure behave in curved spacetime. For traversable WHs, the violation of the null energy constraint near the throat is typically unavoidable, as such violation prevents gravitational collapse and allows the WH to remain open, and therefore, examining the behavior of energy conditions helps identify the regions where exotic matter is required to sustain the WH structure.
In EGR, energy conditions impose constraints on the EMT of the matter fields \cite{Yuennan2025Wormholes,Ditta2025MassiveGravity,yousaf2025interpretation} and the main energy conditions for an anisotropic fluid with energy density $\rho$, radial pressure $P_{rad}$, and tangential pressure $P_{tan}$ are summarized as follows.\\
Weak Energy Condition (WEC): 
\begin{align}
T_{\nu\eta}v^\nu v^\eta \ge 0,\label{b6}
\end{align}
where $v^\nu$ is any timelike vector. For anisotropic matter:
\begin{align}
\rho \ge 0,\qquad \rho + P_{rad} \ge 0,\qquad \rho + P_{tan} \ge 0. \label{b7}
\end{align}
Strong Energy Condition (SEC):
\begin{align}
\left(T_{\nu\eta}-\frac{T}{2}g_{\nu\eta}\right)v^\nu v^\eta \ge 0,\label{b8}
\end{align}
which yields
\begin{align}
\rho + P_{rad} + 2P_{tan} \ge 0,\qquad \rho + P_{rad} \ge 0. \label{b9}
\end{align}
Null Energy Condition (NEC):
\begin{align}
T_{\nu\eta}k^\nu k^\eta \ge 0,\label{b10}
\end{align}
for any null vector $k^\nu$, equivalent to:
\begin{align}
\rho + P_{rad} \ge 0,\qquad \rho + P_{tan} \ge 0. \label{b11}
\end{align}
Dominant Energy Condition (DEC):
\begin{align}
T_{\nu\eta}v^\nu v^\eta \ge 0, \label{b12}
\end{align}
with the anisotropic form:
\begin{align}
\rho \ge 0,\qquad \rho \ge |P_{rad}|,\qquad \rho \ge |P_{tan}|. \label{b13}
\end{align}
These conditions impose strict physical constraints; however, in EGR, constructing a traversable WH typically requires exotic matter that violates the NEC near the throat, whereas for the fuzzy WH configuration studied here, the matter variables $\rho$, $P_{rad}$, and $P_{tan}$ corresponding to the Einasto-density profile together with the constructed shape function are obtained as:
\begin{align}\label{a23rho}
\rho &=\frac{1}{r^2}\left[-\xi ^2+\xi ^2 \left(-\frac{1}{\xi ^2}-\frac{\rho_0 r^2 \sinh \left(\left(\frac{r}{\beta}\right)^{1/\gamma_{1}}\right)}{\xi ^2}+\frac{\rho_0 r^2 \cosh \left(\left(\frac{r}{\beta}\right)^{1/\gamma_{1}}\right)}{\xi ^2}+1\right)+1\right] 
\\\label{a23}
P_ {rad}&=\frac{1}{r^3}\left[\rho_0 \left(\gamma_{1} r^3 \left(\left(\frac{r}{\beta}\right)^{1/\gamma_{1}}\right)^{-3 \gamma_{1}} \Gamma \left(3 \gamma_{1},\left(\frac{r}{\beta}\right)^{1/\gamma_{1}}\right)-\gamma_{1} a_{0}^3 \left(\left(\frac{a_{0}}{\beta}\right){}^{1/\gamma_{1}}\right){}^{-3 \gamma_{1}} \Gamma \left(3 \gamma_{1},\left(\frac{a_{0}}{\beta}\right){}^{1/\gamma_{1}}\right)\right)-a_{0}\right],
\\\nonumber
P_{tan}&=\frac{1}{2} \left[\frac{a_{0}}{r^3}+\rho_0 \left(\frac{\gamma_{1} a_{0}^3 \left(\left(\frac{a_{0}}{\beta}\right){}^{1/\gamma_{1}}\right){}^{-3 \gamma_{1}} \Gamma \left(3 \gamma_{1},\left(\frac{a_{0}}{\beta}\right){}^{1/\gamma_{1}}\right)}{r^3}-e^{-\left(\frac{r}{\beta}\right)^{1/\gamma_{1}}}-\gamma_{1} \left(\left(\frac{r}{\beta}\right)^{1/\gamma_{1}}\right)^{-3 \gamma_{1}} \right.\right.
\\\label{a24}&
\left.\left.\times\Gamma \left(3 \gamma_{1},\left(\frac{r}{\beta}\right)^{1/\gamma_{1}}\right)\right)\right].
\end{align}
To examine the fuzzy WH structure, we focus on the regime $\xi \neq 1$ and Einasto index $\gamma_{1}=3.33$, which controls the shape of the Einasto-density profile model, while the graphical analysis in Fig.~\ref{energydensity} shows the behavior of the density and pressures for $a_0 = 0.6$.
Panel (a) of Fig.~\ref{energydensity} shows that $\rho$ remains positive near the throat and decreases monotonically toward zero, indicating that WH supporting matter is concentrated around $a_0$. In panel (b), the radial pressure $P_{rad}$ is negative near the throat, consistent with the exotic matter requirement, whereas $P_{tan}$ is positive in this region. This anisotropic behavior ($P_{rad} \neq P_{tan}$) reflects the internal stresses necessary to maintain WH stability.
To evaluate the NECs and DECs, we compute:
\begin{align}\nonumber&
\rho+P_ {rad}=\rho_0 \left(-\frac{\gamma_{1} a_{0}^3 \left(\left(\frac{a_{0}}{\beta}\right){}^{1/\gamma_{1}}\right){}^{-3 \gamma_{1}} \Gamma \left(3 \gamma_{1},\left(\frac{a_{0}}{\beta}\right){}^{1/\gamma_{1}}\right)}{r^3}+e^{-\left(\frac{r}{\beta}\right)^{1/\gamma_{1}}}+\gamma_{1} \left(\left(\frac{r}{\beta}\right)^{1/\gamma_{1}}\right)^{-3 \gamma_{1}} \right.
\\\label{7a}&\left.\times\Gamma \left(3 \gamma_{1},\left(\frac{r}{\beta}\right)^{1/\gamma_{1}}\right)\right)-\frac{a_{0}}{r^3}
,
\\\nonumber&
\rho+P_ {tan}=\frac{1}{2} \left(\rho_0 \left(\frac{\gamma_{1} a_{0}^3 \left(\left(\frac{a_{0}}{\beta}\right){}^{1/\gamma_{1}}\right){}^{-3 \gamma_{1}} \Gamma \left(3 \gamma_{1},\left(\frac{a_{0}}{\beta}\right){}^{1/\gamma_{1}}\right)}{r^3}+e^{-\left(\frac{r}{\beta}\right)^{1/\gamma_{1}}}-\gamma_{1} \left(\left(\frac{r}{\beta}\right)^{1/\gamma_{1}}\right)^{-3 \gamma_{1}} \right.\right.
\\\label{7aa}&\left.\left.\times\Gamma \left(3 \gamma_{1},\left(\frac{r}{\beta}\right)^{1/\gamma_{1}}\right)\right)+\frac{a_{0}}{r^3}\right).
\end{align}

\begin{align}\nonumber&
\rho-\left|P_ {rad}\right|=-\left| \frac{\left(\gamma_{1} \Gamma \left(3 \gamma_{1},\left(\frac{a_{0}}{\beta}\right){}^{1/\gamma_{1}}\right) a_{0}^3 \left(\left(\frac{a_{0}}{\beta}\right){}^{1/\gamma_{1}}\right){}^{-3 \gamma_{1}}-r^3 \left(\left(\frac{r}{\beta}\right)^{1/\gamma_{1}}\right)^{-3 \gamma_{1}} \gamma_{1} \Gamma \left(3 \gamma_{1},\left(\frac{r}{\beta}\right)^{1/\gamma_{1}}\right)\right) \rho_0+a_{0}}{r^3}\right|
\\\label{abs1}&
+\rho_0 \left(\cosh \left(\left(\frac{r}{\beta}\right)^{1/\gamma_{1}}\right)-\sinh \left(\left(\frac{r}{\beta}\right)^{1/\gamma_{1}}\right)\right)
,
\\\nonumber& 
\rho-\left|P_ {tan}\right|=-\frac{1}{2} \left| \left(-\gamma_{1} \Gamma \left(3 \gamma_{1},\left(\frac{r}{\beta}\right)^{1/\gamma_{1}}\right) \left(\left(\frac{r}{\beta}\right)^{1/\gamma_{1}}\right)^{-3 \gamma_{1}}+\frac{\left(\left(\frac{a_{0}}{\beta}\right){}^{1/\gamma_{1}}\right){}^{-3 \gamma_{1}} \gamma_{1} \Gamma \left(3 \gamma_{1},\left(\frac{a_{0}}{\beta}\right){}^{1/\gamma_{1}}\right) a_{0}^3}{r^3}
\right.\right.
\\\label{abs2}& 
\left.\left.-\cosh \left(\left(\frac{r}{\beta}\right)^{1/\gamma_{1}}\right)+\sinh \left(\left(\frac{r}{\beta}\right)^{1/\gamma_{1}}\right)\right) \rho_0+\frac{a_{0}}{r^3}\right|
+\rho_0 \left(\cosh \left(\left(\frac{r}{\beta}\right)^{1/\gamma_{1}}\right)-\sinh \left(\left(\frac{r}{\beta}\right)^{1/\gamma_{1}}\right)\right).
\end{align}
The behavior of these quantities is presented in Fig.~\ref{NEC}.
From Fig.~\ref{NEC}(a), it can be observed that radial NEC combination
$\rho+P_{\rm rad}$ becomes negative in vicinity of WH throat, indicating a violation of NEC in radial direction, in contrast, $\rho+P_{\rm tan}$ remains positive for considered parameter choices, showing that NEC is satisfied along tangential direction. Since energy density $\rho$ remains positive in region of interest, WEC exhibits a similar directional behavior: it is violated in radial sector owing to $\rho+P_{\rm rad}<0$, whereas it remains satisfied in tangential sector where $\rho+P_{\rm tan}>0$. This behavior indicates that exotic character of matter supporting traversable WH is mainly associated with radial pressure component and is concentrated near throat. Figure~\ref{NEC}(b) illustrates corresponding behavior of DEC, while quantities $\rho-|P_{\rm rad}|$ as well as $\rho-|P_{\rm tan}|$ show that DEC may be violated or only partially satisfied close to throat, depending on adopted parameter values, however, their behavior improves with increasing radial distance, indicating that influence of exotic stresses becomes progressively weaker away from throat. Consequently, energy condition analysis demonstrates a
localized violation of radial NEC as well as WEC near the WH throat, while
tangential NEC remains satisfied for parameter ranges considered, while this localized exotic behavior is consistent with the matter requirements of a traversable WH geometry.

\section{Stability Analysis of Fuzzy Wormhole Configuration Through TOV Equation, Anisotropy Parameter, and Average Pressure}\label{Stability}

\begin{figure}[h]
\centering
\subfloat[]{{\includegraphics[height=2.3 in, width=3.3 in]{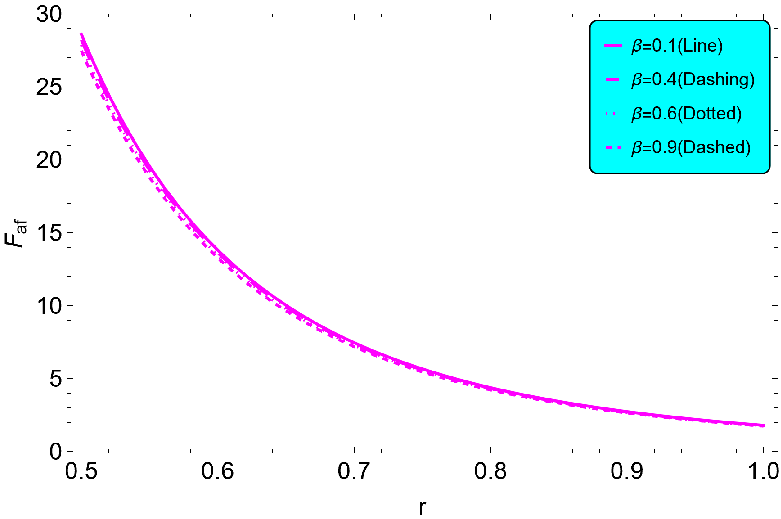}}}
\qquad
\subfloat[]{{\includegraphics[height=2.3 in, width=3.3 in]{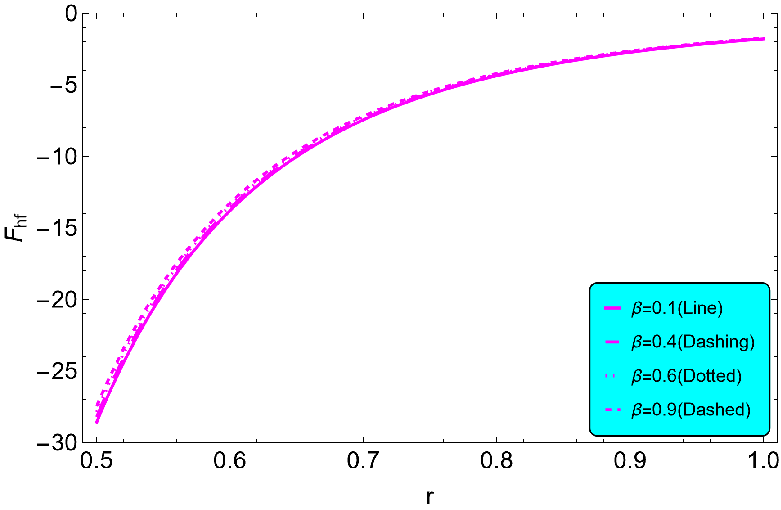}}}
\caption{Plot $F_{af}$ vs $r$ panel (a) and $F_{hf}$ vs $r$ panel (b) for the fuzzy WH geometry at throat radius $a_{0}=0.6$.}
\label{fafh}
\end{figure}

The equilibrium properties of self-gravitating anisotropic configurations can be investigated through several complementary physical criteria, while in the present analysis, we focus on the generalized TOV equation, the anisotropy parameter $\Delta(r)$, and the average pressure $P_{\rm aver}$ to examine the mechanical behavior of the constructed fuzzy WH configuration. These quantities provide useful information regarding the balance of internal forces, the contribution of anisotropic stresses, as well as radial behavior of effective matter supporting WH geometry. In particular, generalized TOV equation describes equilibrium among relevant force components, whereas anisotropy parameter characterizes difference between the radial and tangential pressures. The average pressure further provides information about overall effective pressure distribution within WH spacetime. In gravitational settings, such conservation principles play a central role in determining whether a configuration can resist internal as well as external forces without undergoing collapse or developing instabilities. We study a comprehensive evaluation of stability in this part of work which can be achieved by examining the combined behaviour of the generalized TOV equation, anisotropy parameter $\Delta(r)$, and the average pressure $P_{aver}$. The equilibrium and stability properties of the present anisotropic WH configuration can be examined through the combined behavior of the generalized TOV equation, the anisotropy parameter $\Delta(r)$, and the average pressure $P_{\rm aver}$, in particular, the TOV equation describes balance among the effective force components, whereas sign and radial behavior of $\Delta(r)$ characterize role of anisotropic stresses in supporting the WH geometry. The average pressure provides additional information about the effective matter distribution and should remain finite and regular throughout region of interest. In contrast to conventional stellar configurations, however, a negative average pressure is not necessarily an indication of instability in a traversable WH, since negative effective pressure is naturally associated with the exotic stresses required to sustain throat. In anisotropic configurations, where $P_{\mathrm{rad}} \neq P_{\mathrm{tan}}$, the equilibrium of internal forces becomes more intricate. The generalized TOV equation \cite{albalahi2024isotropization} effectively captures this interplay by accounting for hydrostatic, gravitational, and anisotropic contributions and mathematically it can be presented as 
\begin{equation}\label{GTOV}
\frac{dP_{\mathrm{rad}}}{dr} = -\frac{A_{1}'(r)}{2}\big(\rho + P_{\mathrm{rad}}\big) - \frac{2}{r}\big(P_{\mathrm{rad}} - P_{\mathrm{tan}}\big),
\end{equation}
The graphical behavior of forces for representative parameter sets is displayed in panel (a) and (b) of Fig.~\ref{fafh}, illustrating that the stability criterion is satisfied across a wide range of configurations.
In above mention TOV equation \eqref{GTOV} $A_{1}'(r)$ denotes the derivative of metric potential with respect to radial coordinate $r$, and this expression can be decomposed into three force components:
\begin{description}
  \item[Gravitational Force:] $F_{\mathrm{gf}} = -\frac{A_{1}'(r)}{2}\,(\rho+P_{\mathrm{rad}})$
  \item[Anisotropic Force:] $F_{\mathrm{af}} = -\frac{2}{r}\,(P_{\mathrm{rad}}-P_{\mathrm{tan}})$
  \item[Hydrostatic Force:] $F_{\mathrm{hf}} = -\frac{dP_{\mathrm{rad}}}{dr}$
\end{description}
representing gravitational, anisotropic, and hydrostatic forces, respectively. The equilibrium condition requires
\begin{equation}
\text{Gravitational~Force} + \text{Anisotropic~Force} + \text{Hydrostatic~Force}~\Rightarrow~F_{\mathrm{gf}} + F_{\mathrm{af}} + F_{\mathrm{hf}} = 0.
\end{equation}
In the present scenario, the gravitational force contribution vanishes, reducing the equilibrium condition to
\begin{equation}
\text{Anisotropic~Force} + \text{Hydrostatic~Force}~\Rightarrow~F_{\mathrm{af}} + F_{\mathrm{hf}} = 0.
\end{equation}
A physically viable model requires the radial pressure to be smooth and monotonic behavior, in this case study it ensuring existence of exotic matter.

The anisotropy parameter, defined as
\begin{equation}
\Delta(r) = \text{Tangential Pressure} - \text{Radial Pressure}~\Rightarrow~\Delta(r)=P_{\mathrm{tan}} - P_{\mathrm{rad}},
\end{equation}
quantifies the difference between tangential and radial stresses, while there three possibilities regarding the values of $\Delta$ (i) if positive anisotropy, i.e., ($\Delta>0$), introduces an outward directed force that enhances stability, whereas (ii) if $\Delta<0$ increases the tendency toward gravitational collapse and (iii) if $\Delta=0$ then isotropic behavior appear. To assess the pressure behavior in all spatial directions, the average pressure for the anisotropic fluid is defined as
\begin{equation}
P_{aver} = \frac{\text{Two~Times~of~Tangential~Pressure} + \text{Radial~Pressure}}{3}~\Rightarrow\frac{\big(P_{\mathrm{rad}} + 2P_{\mathrm{tan}}\big)}{3},
\end{equation}
which is an essential quantity for analyzing energy conditions, causality bounds, and the stiffness of the matter distribution.
Using the shape function derived from the Einasto-density profile model, the forces $F_{\mathrm{af}}$ and $F_{\mathrm{hf}}$, the anisotropy parameter and average pressure are obtained as
\begin{align}\nonumber
F_{\mathrm{af}} &= \frac{1}{r^4}\left[\rho_0 \left(3\gamma_{1} a_0^3 \Big[\left(\frac{a_0}{\beta}\right)^{\left\{\frac{1}{\gamma_{1}}\right\}}\Big]^{-3\gamma_{1}} \Gamma \left(3\gamma_{1},\left(\frac{a_0}{\beta}\right)^{\left\{\frac{1}{\gamma_{1}}\right\}}\right) + r^3 \left[-e^{-\left\{\frac{r}{\beta}\right\}^{\left\{\frac{1}{\gamma_{1}}\right\}}} - 3\gamma_{1}\left\{\frac{r}{\beta}\right\}^{-3\gamma_{1}} \right.\right.\right.
\\\label{a33}&\left.\left.\left.\times\Gamma\left(3\gamma_{1},\left\{\frac{r}{\beta}\right\}^{\left\{\frac{1}{\gamma_{1}}\right\}}\right)\right]\right) + 3 a_0\right],
\end{align}
\begin{align}\nonumber
F_{\mathrm{hf}} &= \frac{1}{r^3}\left[\frac{3\gamma_{1}\rho_0 r^3 \Big(\frac{r}{\beta}\Big)^{-3\gamma_{1}-1} \Big(\frac{r}{\beta}\Big)^{\frac{1}{\gamma_{1}}-1} \Gamma\left(3\gamma_{1},\left\{\frac{r}{\beta}\right\}^{\left\{\frac{1}{\gamma_{1}}\right\}}\right)}{\beta} + \rho_0 r^2 e^{-\left\{\frac{r}{\beta}\right\}^{\left\{\frac{1}{\gamma_{1}}\right\}}} - 3\gamma_{1}\rho_0 r^2 \Big(\frac{r}{\beta}\Big)^{-3\gamma_{1}} 
\right.
\\\nonumber&\left.\times\Gamma\left(3\gamma_{1},\left\{\frac{r}{\beta}\right\}^{\left\{\frac{1}{\gamma_{1}}\right\}}\right)\right] - \frac{3}{r^4}\left(a_0 \Big(\big(\frac{a_0}{\beta}\big)^{\left\{\frac{1}{\gamma_{1}}\right\}}\Big)^{-3\gamma_{1}}\big(\gamma_{1}\rho_0 a_0^2 \Gamma(3\gamma_{1},\left\{\frac{a_0}{\beta}\right\}^{\left\{\frac{1}{\gamma_{1}}\right\}}) + (\frac{a_0}{\beta})^{3\gamma_{1}}\big)\right.\\ \label{a34}
&\left.\xi^2(-r)+r+\,\gamma_{1}\rho_0(-r^3)\left\{\frac{r}{\beta}\right\}^{-3\gamma_{1}} \frac{\Gamma(3\gamma_{1},\left\{\frac{r}{\beta}\right\}^{\left\{\frac{1}{\gamma_{1}}\right\}})+\xi^2 r - r}{\xi^2}\right).
\end{align}
\begin{align}\nonumber
\Delta &= \frac{1}{2}\left[\rho_0\left(\frac{3\gamma_{1} a_0^3 \left(\left(\frac{a_0}{\beta}\right)^{\left\{\frac{1}{\gamma_{1}}\right\}}\right)^{-3\gamma_{1}} \Gamma\left(3\gamma_{1},\left(\frac{a_0}{\beta}\right)^{\left\{\frac{1}{\gamma_{1}}\right\}}\right)}{r^3}-e^{-\left\{\frac{r}{\beta}\right\}^{\left\{\frac{1}{\gamma_{1}}\right\}}} - 3\gamma_{1}\left\{\frac{r}{\beta}\right\}^{-3\gamma_{1}}\right.\right.\\\label{anisotropy}
&\left.\left. \Gamma\left(3\gamma_{1},\left\{\frac{r}{\beta}\right\}^{\left\{\frac{1}{\gamma_{1}}\right\}}\right)\right) + \frac{3a_0}{r^3}\right].
\end{align}
\begin{align}\nonumber
P_{aver} &= \frac{1}{3}\left[\frac{\rho_0 \left(n r^3\left\{\frac{r}{\beta}\right\}^{-3\gamma_{1}} \Gamma\left(3\gamma_{1},\left\{\frac{r}{\beta}\right\}^{\left\{\frac{1}{\gamma_{1}}\right\}}\right)- n a_0^3\left(\left(\frac{a_0}{\beta}\right)^{\left\{\frac{1}{\gamma_{1}}\right\}}\right)^{-3\gamma_{1}}\Gamma\left(3\gamma_{1},\left(\frac{a_0}{\beta}\right)^{\left\{\frac{1}{\gamma_{1}}\right\}}\right)\right)-a_0}{r^3} + \frac{a_0}{r^3}\right.\\ \label{paverage}
& \left.+\,\rho_0\left(\frac{n a_0^3\left(\left(\frac{a_0}{\beta}\right)^{\left\{\frac{1}{\gamma_{1}}\right\}}\right)^{-3\gamma_{1}}\Gamma\left(3\gamma_{1},\left(\frac{a_0}{\beta}\right)^{\left\{\frac{1}{\gamma_{1}}\right\}}\right)}{r^3}-e^{-\left\{\frac{r}{\beta}\right\}^{\left\{\frac{1}{\gamma_{1}}\right\}}} - n\left\{\frac{r}{\beta}\right\}^{-3\gamma_{1}}\Gamma\left(3\gamma_{1},\left\{\frac{r}{\beta}\right\}^{\left\{\frac{1}{\gamma_{1}}\right\}}\right)\right)\right].
\end{align}

\begin{figure}[h]
\centering
\subfloat[]{{\includegraphics[height=2.8 in, width=3.4 in]{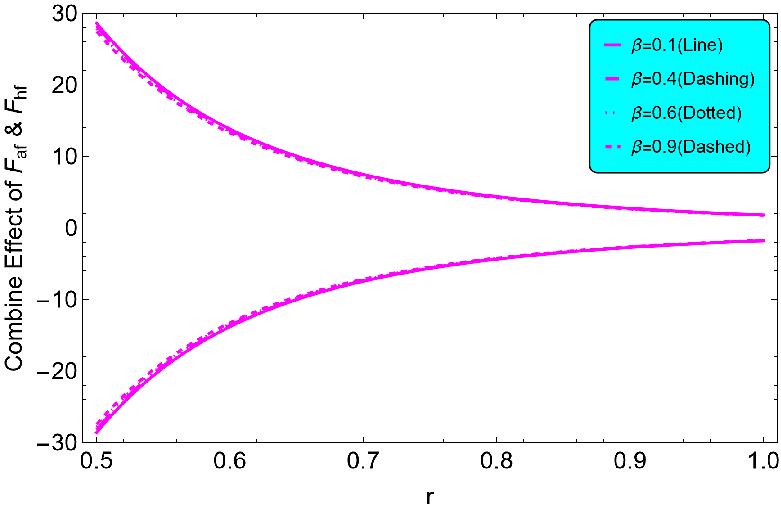}}}
\qquad
\subfloat[]{{\includegraphics[height=2.8 in, width=3.4 in]{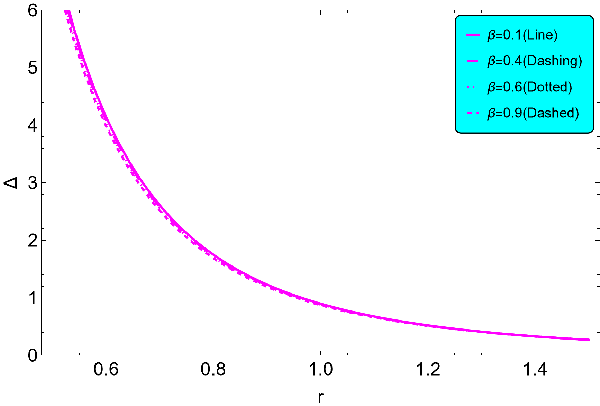}}}
\qquad
\subfloat[]{{\includegraphics[height=2.8 in, width=3.4 in]{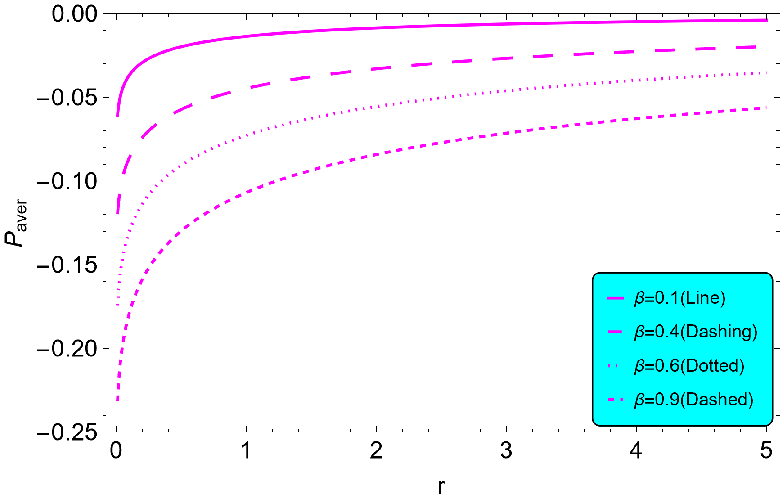}}}
\caption{Pictorial representation of equilibrium forces panel (a), anisotropy parameter panel (b) and average pressure panel (c) at $a_0=0.6$ and $\gamma_{1}=3.33$ and varying Einasto-density profile model parameters.}\label{TOVAnAP}
\end{figure}
From Fig. \ref{TOVAnAP}, the following features can be observed:
\begin{itemize}
\item[(a)] Figure~\ref{TOVAnAP}(a) shows that the anisotropic force $F_{\rm af}$ and hydrostatic force $F_{\rm hf}$ act in opposite directions and effectively counterbalance each other. Since the gravitational contribution vanishes for the adopted constant redshift function, the condition $F_{\rm af}+F_{\rm hf}=0$ provides the required force balance for the static WH configuration.
\item[(b)] Figure~\ref{TOVAnAP}(b) demonstrates that the anisotropy parameter $\Delta(r)=P_{\rm tan}-P_{\rm rad}$ remains positive for the considered choices of the Einasto profile and monopole parameters, therefore, the anisotropic stress generates an outward directed contribution, which assists in preventing the WH throat from undergoing gravitational collapse.
\item[(c)] Figure~\ref{TOVAnAP}(c) illustrates that the average pressure
$P_{\rm aver}=(P_{\rm rad}+2P_{\rm tan})/3$ remains finite and exhibits a
regular radial behavior, while taking negative values over the considered
domain, such negative average pressure should not be interpreted as a
conventional stellar instability criterion, instead, it reflects exotic
effective stresses associated with matter threading traversable WH, consistent with negative radial pressure as well as violation of radial
NEC found near the throat.
\end{itemize}
Consequently, combined behavior of equilibrium forces as well as anisotropy parameter indicates that proposed fuzzy WH configuration can
maintain mechanical equilibrium for adopted parameter choices, while the negative average pressure further characterizes exotic nature of supporting matter rather than contradicting equilibrium of WH
geometry. Thus, the TOV force balance, together with the outward directed
anisotropic stresses, provides evidence for equilibrium behavior of the
Einasto dark matter supported WH configuration within Einstein gravity.

\section{Formalism of thin-shell around WH by using fuzzy BH}\label{TSW}

Here, we are interested to develop the geometrical structure thin-shell around WH by considering the outer manifold as a fuzzy BH described in \cite{batic2021fuzzy} and the inner manifold presented by calculated traversable WH. We would like to understand the effect of the shape functions and the characteristics of the linearized radial perturbations on the physically realistic thin-shell stable configurations. Mathematically, the line element of outer BH structure can be presented as \cite{batic2021fuzzy}
\begin{equation}\label{1aa}
ds^{2}_+=-\Pi_+(r_+)^{-1}dr^{2}_+-r^{2}_+
d\theta^{2}_+-r^{2}_+\sin^2{\theta}_+
d\phi^{2}_++\Pi_+(r_+)dt^{2}_+,
\end{equation}
where
\begin{equation}\label{2aa}
\Pi_+(r_+)=1-\frac{(2 \mathcal{U}) \left(\frac{r}{h}\right)^2}{\Gamma \left(3 \gamma _1+1\right)},
\end{equation}
here $\mathcal{U}=m/h$ where total mass $m$ associated
to the Einasto density profile and $h$ is the scale length \cite{batic2021fuzzy}. Further, the inner geometrical structure (-) is considered as a WH structure. The respective line element is given as 
\begin{equation}\label{2aaa}
ds^2_-=-e^{2\varphi(r_-)}dt^2_-+r^2_-d\theta^2_-+\frac{dr^2_-}{\Pi_-(r_-)}+r^2_-sin^2\theta_-
d\phi^2_-,
\end{equation}
where
\begin{equation}\label{2aaaa}
\Pi_-(r_-)=\xi ^2\left(-\frac{B(r_-)}{r_-}+1\right).
\end{equation}

Visser \cite{visser1989traversable} came up with a method of cut and paste of constructing a thin shell by joining together two identical copies of BH spacetimes at a hypersurface. This method is adopted in this paper to construct the geometry of the thin-shell around the WH spacetime. In this regard, we divide this spacetime into two regions given as
\begin{eqnarray}\label{3aa}
\mathcal{M}^{-}=\left\lbrace r^{-}\leq \mathcal{X},r_{h}<\mathcal{X}\right\rbrace, \quad  \mathcal{M}^{+}=\left\lbrace r^{+}\leq \mathcal{X},r_{h}<\mathcal{X}\right\rbrace,
\end{eqnarray}
where $r_h$ is the considered BH horizon radius. Also, $\mathcal{X}$ is shell radius which must be greater than the horizon structure. The hypersurface plays a remarkable role to join the inner and outer manifolds. It can be defined as $
\Sigma=\left\lbrace r^{\pm}=\mathcal{X},\mathcal{X}>r_{h}\right\rbrace.
$ 
This formalism develops manifold as $\mathcal{M} = \mathcal{M}^{-} \cup \mathcal{M}^{+}$. It does not have any event horizon as well as singularity. For outer and inner manifolds the coordinates can be denoted as $y_{\pm }^{\gamma} = (t_{\pm}, r_{\pm}, \theta_{\pm}, \phi_{\pm})$. For, hypersurface, we can write $\eta^{i} = (\tau, \theta, \phi)$. The term $\tau$ represented the proper time. We can related these coordinates by using the following transformation
\begin{eqnarray}\label{5aa}
g_{ij}=\frac{\partial y^{\gamma}}{\partial\eta^{i}}\frac{\partial
y^{\xi}}{\partial\eta^{j}}g_{\gamma\xi}.
\end{eqnarray}
For hypersurface, we can define 
\begin{eqnarray}\nonumber
\Sigma:R(r,\tau)=r-\mathcal{X}(\tau)=0.
\end{eqnarray}
The respective Lanczos equations can be written as 
\begin{equation}\label{6aa}
S_{\xi}^{\alpha}=\frac{1}{8\pi}(\delta_{\xi}^{\alpha}
\zeta_{\gamma}^{\gamma}-\zeta_{\xi}^{\alpha}),
\end{equation}
where $\zeta_{\alpha\xi}=K^{+}_{\alpha\xi}-K^{-}_{\alpha\xi}$
and $K^{-}_{\alpha\xi}$ express the extrinsic
curvature components. For the case perfect fluid distribution, we write ${S^{\alpha}}_{\xi}=diag(\mathcal{Z},p,p)$. On the hyper-surface \(\Sigma\), the surface density and surface pressure are denoted \(\mathcal{Z}\) and \(p\), respectively. The extrinsic curvatures can be defined as
\begin{equation}\label{7aa}
{K_{\alpha\xi}^{\pm}}= -n_{\mu}^\pm \left[\frac{\partial^2
y^{\mu}_\pm}{\partial \eta^{\alpha}
\eta^{\xi}}+\Gamma^{\mu}_{\lambda\nu}\left(\frac{\partial
y^{\lambda}_\pm}{\partial\eta^{\alpha}}\right)\left(\frac{\partial
y^{\nu}_\pm}{\partial\eta^{\xi}}\right)\right],
\end{equation}
where unit normal is denoted by $n_{\pm}^{\mu}$. Further, we can obtain
\begin{eqnarray}\label{9aa}
\mathcal{Z}&=&-\frac{[K^\theta_\theta]}{4\pi}=-\frac{\sqrt{\dot{\mathcal{X}}^2-\frac{(2 \mathcal{U}) \left(\frac{\mathcal{X}}{h}\right)^2}{\Gamma \left(3 \gamma _1+1\right)}+1}-\sqrt{\frac{1-\frac{B(\mathcal{X})}{\mathcal{X}}}{\xi ^2}+\dot{\mathcal{X}}^2}}{4 \pi  \mathcal{X}},
\\\label{10aa}
p&=&\frac{[K^\theta_\theta]+[K_{\tau}^{\tau}]}{8\pi}=\frac{\frac{2 (\mathcal{X}+1) B(\mathcal{X})-\mathcal{X} \left(\mathcal{X} \left(-B''(\mathcal{X})+2 \dot{\mathcal{X}}^2 \xi ^2+2 \ddot{\mathcal{X}} \mathcal{X} \xi ^2+2\right)+2 B'(\mathcal{X})\right)}{\mathcal{X}^2 \xi ^2 \sqrt{-\frac{B(\mathcal{X})}{\mathcal{X} \xi ^2}+\dot{\mathcal{X}}^2+\frac{1}{\xi ^2}}}+\frac{2 h^2 \Gamma \left(3 \gamma _1+1\right) \left(\dot{\mathcal{X}}^2+\ddot{\mathcal{X}} \mathcal{X}+1\right)-8 \mathcal{X}^2 \mathcal{U}}{h^2 \Gamma \left(3 \gamma _1+1\right) \sqrt{\dot{\mathcal{X}}^2-\frac{2 \mathcal{X}^2 \mathcal{U}}{h^2 \Gamma \left(3 \gamma _1+1\right)}+1}}}{8 \pi  \mathcal{X}},
\end{eqnarray}
while
\begin{eqnarray}\nonumber
4 \pi  \mathcal{X}(\mathcal{Z}+2p)&=&\frac{2 (\mathcal{X}+1) B(\mathcal{X})-\mathcal{X} \left(\mathcal{X} \left(-B''(\mathcal{X})+2 \dot{\mathcal{X}}^2 \xi ^2+2 \ddot{\mathcal{X}} \mathcal{X} \xi ^2+2\right)+2 B'(\mathcal{X})\right)}{\mathcal{X}^2 \xi ^2 \sqrt{-\frac{B(\mathcal{X})}{\mathcal{X} \xi ^2}+\dot{\mathcal{X}}^2+\frac{1}{\xi ^2}}}+\sqrt{-\frac{B(\mathcal{X})}{\mathcal{X} \xi ^2}+\dot{\mathcal{X}}^2+\frac{1}{\xi ^2}}\\\label{11aa}&+&\frac{2 h^2 \Gamma \left(3 \gamma _1+1\right) \left(\dot{\mathcal{X}}^2+\ddot{\mathcal{X}} \mathcal{X}+1\right)-8 \mathcal{X}^2 \mathcal{U}}{h^2 \Gamma \left(3 \gamma _1+1\right) \sqrt{\dot{\mathcal{X}}^2-\frac{2 \mathcal{X}^2 \mathcal{U}}{h^2 \Gamma \left(3 \gamma _1+1\right)}+1}}-\sqrt{\dot{\mathcal{X}}^2-\frac{2 \mathcal{X}^2 \mathcal{U}}{h^2 \Gamma \left(3 \gamma _1+1\right)}+1}.
\end{eqnarray}
where $\dot{\mathcal{X}}=d\mathcal{X}/d\tau$. At $\mathcal{X}=\mathcal{X}_0$, $\dot{\mathcal{X}_0}=0=\ddot{\mathcal{X}_0}$, which shows there no motion at equilibrium shell radius. Hence, we can write
\begin{eqnarray}\label{12aa}
\mathcal{Z}_0=-\frac{\sqrt{\frac{(2 \mathcal{U}) \left(\frac{\mathcal{X}_0}{h}\right)^2}{\Gamma \left(3 \gamma _1+1\right)}+1}-\sqrt{\frac{1-\frac{B(\mathcal{X}_0)}{\mathcal{X}_0}}{\xi ^2}}}{4 \pi  \mathcal{X}_0},\quad
p_0=\frac{\frac{\mathcal{X}_0 \left(\mathcal{X}_0 B''(\mathcal{X}_0)-2 \left(B'(\mathcal{X}_0)+\mathcal{X}_0\right)\right)+2 (\mathcal{X}_0+1) B(\mathcal{X}_0)}{\mathcal{X}_0^2 \xi ^2 \sqrt{\frac{\mathcal{X}_0-B(\mathcal{X}_0)}{\mathcal{X}_0 \xi ^2}}}+\frac{2 \left(h^2 \Gamma \left(3 \gamma _1+1\right)-4 \mathcal{X}_0^2 \mathcal{U}\right)}{h^2 \Gamma \left(3 \gamma _1+1\right) \sqrt{1-\frac{2 \mathcal{X}_0^2 \mathcal{U}}{h^2 \Gamma \left(3 \gamma _1+1\right)}}}}{8 \pi  \mathcal{X}_0},
\end{eqnarray}
and
\begin{equation}\label{13aa}
\mathcal{Z}_0+2p_0=\frac{\frac{\mathcal{X}_0 \left(\mathcal{X}_0 B''(\mathcal{X}_0)-2 \left(B'(\mathcal{X}_0)+\mathcal{X}_0\right)\right)+2 (\mathcal{X}_0+1) B(\mathcal{X}_0)}{\mathcal{X}_0^2 \xi ^2 \sqrt{\frac{\mathcal{X}_0-B(\mathcal{X}_0)}{\mathcal{X}_0 \xi ^2}}}+\sqrt{\frac{\mathcal{X}_0-B(\mathcal{X}_0)}{\mathcal{X}_0 \xi ^2}}+\frac{2 \left(h^2 \Gamma \left(3 \gamma _1+1\right)-4 \mathcal{X}_0^2 \mathcal{U}\right)}{h^2 \Gamma \left(3 \gamma _1+1\right) \sqrt{1-\frac{2 \mathcal{X}_0^2 \mathcal{U}}{h^2 \Gamma \left(3 \gamma _1+1\right)}}}-\sqrt{1-\frac{2 \mathcal{X}_0^2 \mathcal{U}}{h^2 \Gamma \left(3 \gamma _1+1\right)}}}{4 \pi  \mathcal{X}_0}.
\end{equation}
These equations are usefull to determine the shell energy density and pressure at $\mathcal{X}=\mathcal{X}_0$.

\section{Thin-shell Stability}\label{Tinshel_stability}

In order to observe the stable configuration of the shell around WH configuration in the background of fuzzy BH solution, we perturb the effective potential about $\mathcal{X}=\mathcal{X}_0$. for this purpose, we can determine effective potential by taking  Eq.(\ref{9aa}) as:
\begin{equation}\label{14aa}
\dot{\mathcal{X}}^2+V(\mathcal{X})=0.
\end{equation}
where
\begin{equation}\label{15aa}
V(\mathcal{X})=-\frac{\frac{4 \mathcal{X}^3 \mathcal{U} \left(-B(\mathcal{X})+\mathcal{X} \xi ^2 \left(16 \pi ^2 \mathcal{X}^2 \mathcal{Z} ^2-1\right)+\mathcal{X}\right)}{h^2 \xi ^2 \Gamma \left(3 \gamma _1+1\right)}+\frac{\left(B(\mathcal{X})+\mathcal{X} \left(\xi ^2 (1-4 \pi  \mathcal{X} \mathcal{Z} )^2-1\right)\right) \left(B(\mathcal{X})+\mathcal{X} \left((4 \pi  \mathcal{X} \mathcal{Z}  \xi +\xi )^2-1\right)\right)}{\xi ^4}+\frac{4 \mathcal{X}^6 \mathcal{U}^2}{h^4 \Gamma \left(3 \gamma _1+1\right){}^2}}{64 \pi ^2 \mathcal{X}^4 \mathcal{Z} ^2}.
\end{equation}
\begin{figure*}
\centering
\includegraphics[width = 8.5cm,height=6.5cm]{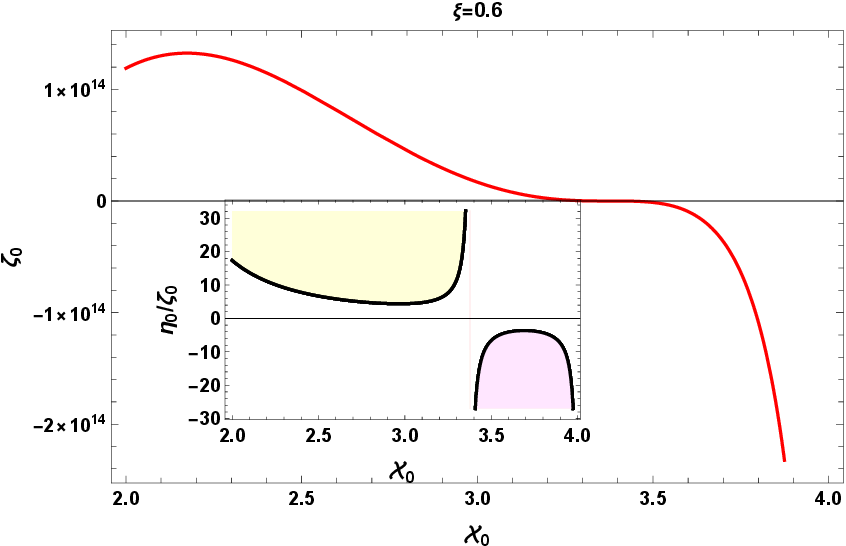}~~~~~~~~\includegraphics[width = 8cm,height=6.5cm]{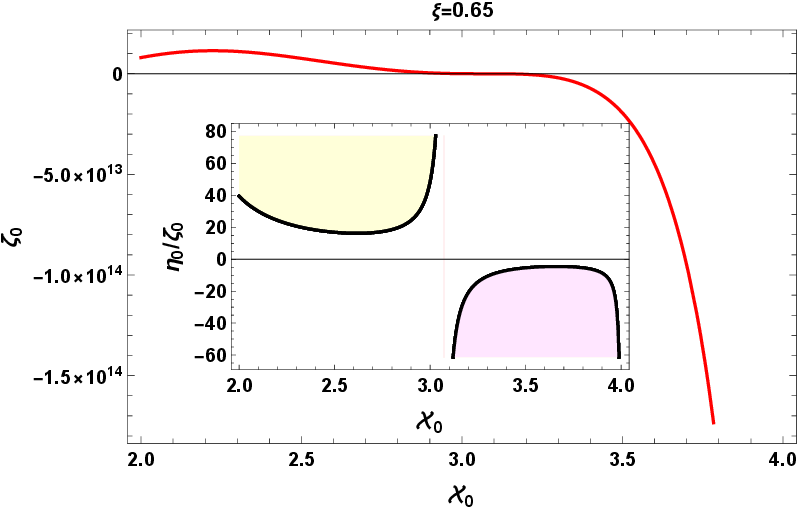}
\caption{\label{p1} Plot of $\zeta_0$ (red line) and $\eta_0/\zeta_0$ (regional plot). The colored regions expressed the stability and unshaded regions denote the unstable configurations.}
\includegraphics[width = 8cm,height=6.5cm]{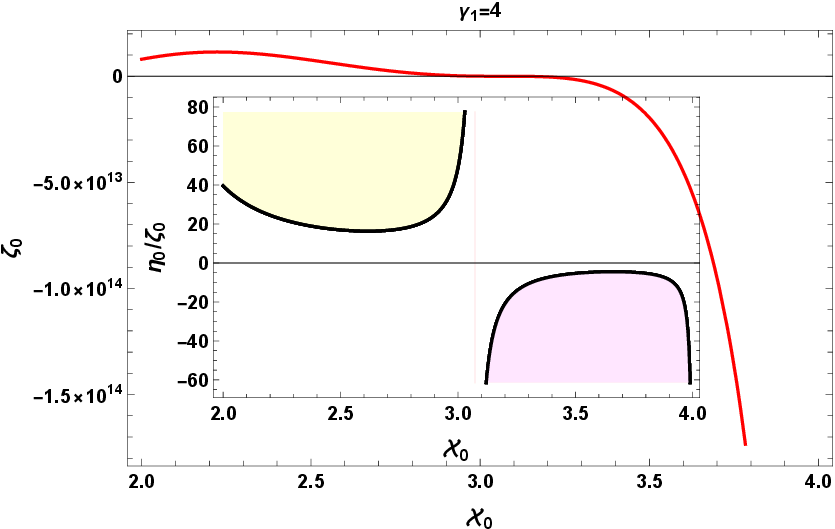}~~~~~~~\includegraphics[width = 8cm,height=6.5cm]{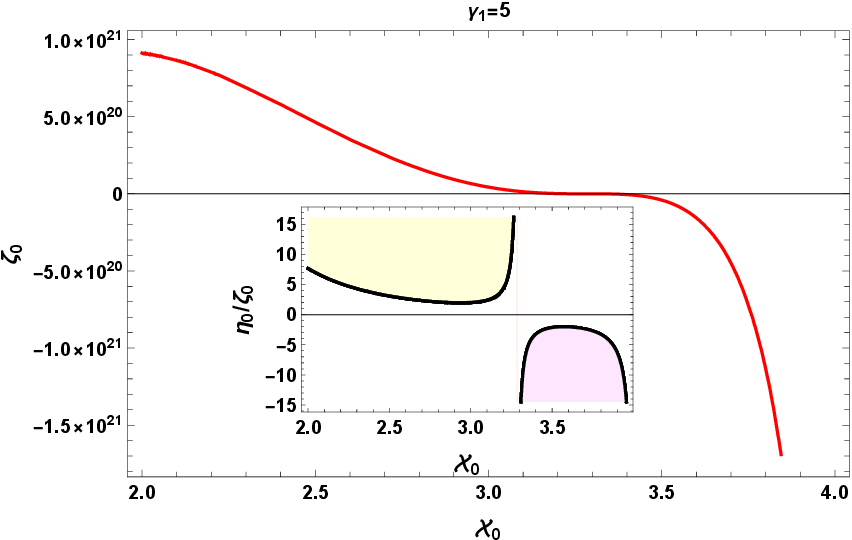}
\caption{\label{p2}Plot of $\zeta_0$ (red line) and $\eta_0/\zeta_0$ (regional plot).}
\includegraphics[width = 8cm,height=6.5cm]{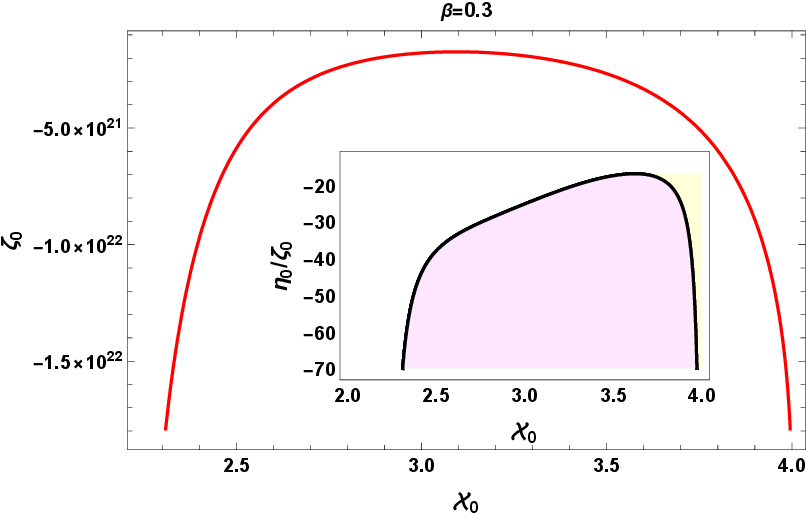}~~~~~~~~\includegraphics[width = 8cm,height=6.5cm]{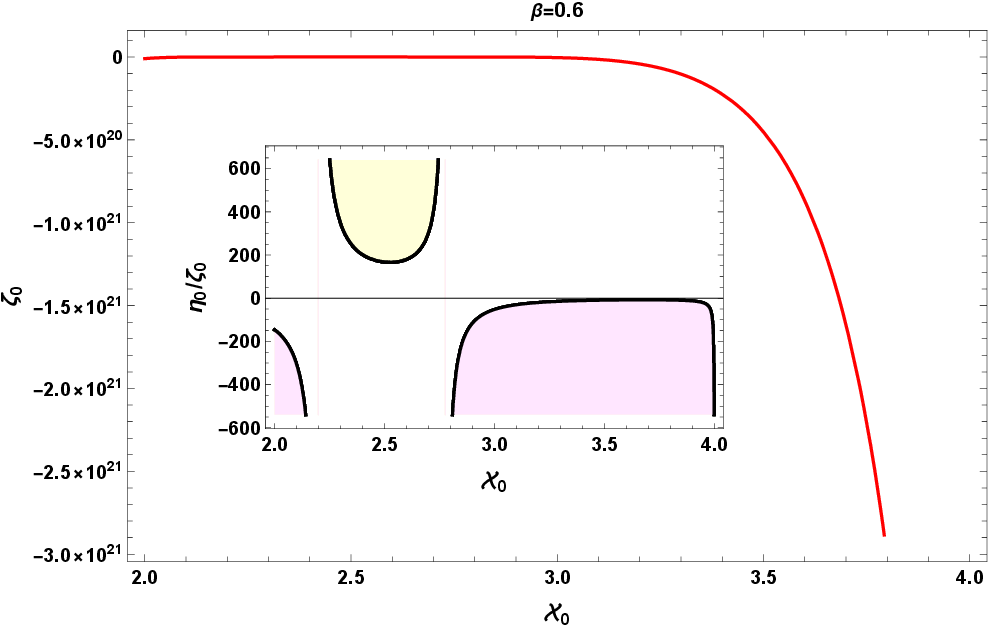}
\caption{\label{p3} Plot of $\zeta_0$ (red line) and $\eta_0/\zeta_0$ (regional plot).}
\end{figure*}

\begin{figure*}
\centering
\includegraphics[width = 8.5cm,height=6.5cm]{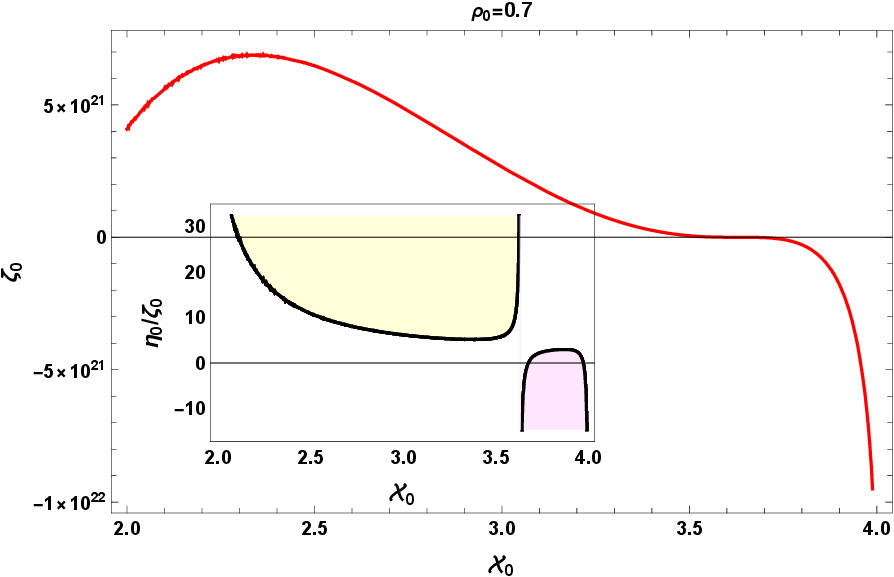}~~~~~~~~\includegraphics[width = 8cm,height=6.5cm]{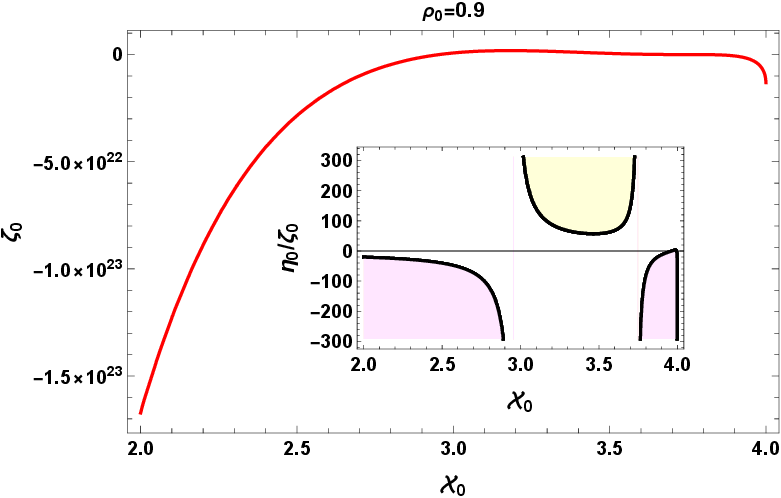}
\caption{\label{p4}Plot of $\zeta_0$ (red line) and $\eta_0/\zeta_0$ (regional plot).}
\end{figure*}
The respective energy conservation constraints can be expressed as
\begin{equation}\label{15b}
p \frac{d}{d\tau}(4\pi \mathcal{X}^2)+\frac{d}{d\tau}(4\pi \mathcal{X}^2\mathcal{Z})=0,
\end{equation}
it yields
\begin{equation}\label{15bb}
\mathcal{Z}'=-\frac{2(\mathcal{Z}+p(\mathcal{Z}))}{\mathcal{X}}.
\end{equation}
By perturbing the effective potential about $\mathcal{X}=\mathcal{X}_0$, we get
\begin{equation}\label{15bbb}
V(\mathcal{X})=V(\mathcal{X}_0)+(\mathcal{X}-\mathcal{X}_0)V'(\mathcal{X}_0)+\frac{1}{2}
(\mathcal{X}-\mathcal{X}_0)^2V''(\mathcal{X}_0)+O[(\mathcal{X}-\mathcal{X}_0)^3].
\end{equation}
We get $V(\mathcal{X}_0)=0=V'(\mathcal{X}_0)$. Therefore, we can write
\begin{equation}\label{16aa}
V(\mathcal{X})=\frac{1}{2}(-\mathcal{X}_0+\mathcal{X})^2V''(\mathcal{X}_{0}).
\end{equation}
Further, the shell's mass can be written as
\begin{eqnarray}\nonumber M(\mathcal{X}_0)=4\pi \mathcal{X}^2_0 \mathcal{Z}_0, \quad \quad M'(\mathcal{X}_0)=-8\pi \mathcal{X}_0 p_0,\quad
M''(\mathcal{X}_0)=-8\pi p_0+16\pi \lambda_0^2 (\mathcal{Z}_0+p_0),
\end{eqnarray}
here the EoS parameter can be written as $\lambda_0^2=dp/d\mathcal{Z}|_{\mathcal{X}=\mathcal{X}_0}$, which yields
\begin{eqnarray}\nonumber
V''(\mathcal{X}_0)&=&-\left(-\mathcal{X}_0^4 M(\mathcal{X}_0) \left(2 \mathcal{X}_0^3 \mathcal{U} \xi ^2-h^2 \Gamma \left(3 \gamma _1+1\right) \left(B(\mathcal{X}_0)+\mathcal{X}_0 \left(\xi ^2-1\right)\right)\right) \left(h^2 \Gamma \left(3 \gamma _1+1\right) \left(-4 M'(\mathcal{X}_0) \right.\right.\right.\\\nonumber&\times&\left.\left.\left.\left(B'(\mathcal{X}_0)+\xi ^2-1\right)-\left(B(\mathcal{X}_0)+\mathcal{X}_0 \left(\xi ^2-1\right)\right) M''(\mathcal{X}_0)\right)+2 \mathcal{X}_0^3 \mathcal{U} \xi ^2 M''(\mathcal{X}_0)+24 \mathcal{X}_0^2 \mathcal{U} \xi ^2 M'(\mathcal{X}_0)\right)\right.\\\nonumber&+&\left.h^2 \mathcal{X}_0 \xi ^2 \Gamma \left(3 \gamma _1+1\right) M(\mathcal{X}_0)^4 \left(h^2 \Gamma \left(3 \gamma _1+1\right) \left(\mathcal{X}_0 \left(\mathcal{X}_0 B''(\mathcal{X}_0)-2 B'(\mathcal{X}_0)+\xi ^2 M'(\mathcal{X}_0)^2\right)+2 B(\mathcal{X}_0)\right)\right.\right.\\\nonumber&+&\left.\left.4 \mathcal{X}_0^3 \mathcal{U} \xi ^2\right)+\mathcal{X}_0^4 M(\mathcal{X}_0)^2 \left(h^4 \Gamma \left(3 \gamma _1+1\right){}^2 \left(\left(B(\mathcal{X}_0)+\mathcal{X}_0 \left(\xi ^2-1\right)\right) B''(\mathcal{X}_0)+\left(B'(\mathcal{X}_0)+\xi ^2-1\right)^2\right)\right.\right.\\\nonumber&-&\left.\left.2 h^2 \mathcal{X}_0 \mathcal{U} \xi ^2 \Gamma \left(3 \gamma _1+1\right) \left(\mathcal{X}_0 \left(\mathcal{X}_0 B''(\mathcal{X}_0)+6 \left(B'(\mathcal{X}_0)+2 \xi ^2-2\right)\right)+6 B(\mathcal{X}_0)\right)+60 \mathcal{X}_0^4 \mathcal{U}^2 \xi ^4\right)\right.\\\nonumber&+&\left.3 \mathcal{X}_0^4 M'(\mathcal{X}_0)^2 \left(h^2 \Gamma \left(3 \gamma _1+1\right) \left(B(\mathcal{X}_0)+\mathcal{X}_0 \left(\xi ^2-1\right)\right)-2 \mathcal{X}_0^3 \mathcal{U} \xi ^2\right){}^2+h^4 \mathcal{X}_0 \xi ^4 \Gamma \left(3 \gamma _1+1\right){}^2 M(\mathcal{X}_0)^5 \right.\\\label{17aaa}&\times&\left.\left(\mathcal{X}_0 M''(\mathcal{X}_0)-4 M'(\mathcal{X}_0)\right)+3 h^4 \xi ^4 \Gamma \left(3 \gamma _1+1\right){}^2 M(\mathcal{X}_0)^6\right)\left(2 h^4 \mathcal{X}_0^4 \xi ^4 \Gamma \left(3 \gamma _1+1\right){}^2 M(\mathcal{X}_0)^4\right)^{-1}.
\end{eqnarray}

By using the stability constraints as $V''(\mathcal{X}_0)>0$, we get
\begin{eqnarray}\nonumber
&-&\left(2 \mathcal{Z}  \left(2 \mathcal{X}_0^3 \mathcal{U} \xi ^2-h^2 \Gamma \left(3 \gamma _1+1\right) \left(B(\mathcal{X}_0)+\mathcal{X}_0 \left(\xi ^2-1\right)\right)\right) \left(-4 h^2 \mathcal{X}_0 p \Gamma \left(3 \gamma _1+1\right) \left(B'(\mathcal{X}_0)+\xi ^2-1\right)+h^2 \Gamma \left(3 \gamma _1+1\right) \right.\right.\\\nonumber&\times&\left.\left.\left(B(\mathcal{X}_0)+\mathcal{X}_0 \left(\xi ^2-1\right)\right) ((2 H-1) p+2 H \mathcal{Z} )-4 H \mathcal{X}_0^3 \mathcal{U} \xi ^2 (p+\mathcal{Z} )+26 \mathcal{X}_0^3 p \mathcal{U} \xi ^2\right)+16 \pi ^2 h^2 \mathcal{X}_0^3 \mathcal{Z} ^4 \xi ^2 \Gamma \left(3 \gamma _1+1\right) \right.\\\nonumber&\times&\left.\left(h^2 \Gamma \left(3 \gamma _1+1\right) \left(\mathcal{X}_0 \left(\mathcal{X}_0 \left(B''(\mathcal{X}_0)+64 \pi ^2 \mathcal{X}_0 p^2 \xi ^2\right)-2 B'(\mathcal{X}_0)\right)+2 B(\mathcal{X}_0)\right)+4 \mathcal{X}_0^3 \mathcal{U} \xi ^2\right)+\mathcal{X}_0^2 \mathcal{Z} ^2 \left(h^4 \Gamma \left(3 \gamma _1+1\right){}^2 \right.\right.\\\nonumber&\times&\left.\left.\left(\left(B(\mathcal{X}_0)+\mathcal{X}_0 \left(\xi ^2-1\right)\right) B''(\mathcal{X}_0)+\left(B'(\mathcal{X}_0)+\xi ^2-1\right)^2\right)-2 h^2 \mathcal{X}_0 \mathcal{U} \xi ^2 \Gamma \left(3 \gamma _1+1\right) \left(\mathcal{X}_0 \left(\mathcal{X}_0 B''(\mathcal{X}_0)\right.\right.\right.\right.\\\nonumber&+&\left.\left.\left.\left.6 \left(B'(\mathcal{X}_0)+2 \xi ^2-2\right)\right)+6 B(\mathcal{X}_0)\right)+60 \mathcal{X}_0^4 \mathcal{U}^2 \xi ^4\right)+12 \left(h^2 p \Gamma \left(3 \gamma _1+1\right) \left(B(\mathcal{X}_0)+\mathcal{X}_0 \left(\xi ^2-1\right)\right)-2 \mathcal{X}_0^3 p \mathcal{U} \xi ^2\right){}^2\right.\\\nonumber&+&\left.512 \pi ^4 h^4 \mathcal{X}_0^6 \mathcal{Z} ^5 \xi ^4 \Gamma \left(3 \gamma _1+1\right){}^2 ((2 H+3) p+2 H \mathcal{Z} )+768 \pi ^4 h^4 \mathcal{X}_0^6 \mathcal{Z} ^6 \xi ^4 \Gamma \left(3 \gamma _1+1\right){}^2\right)\\\label{18ab}&\times&\left(32 \pi ^2 h^4 \mathcal{X}_0^6 \mathcal{Z} ^4 \xi ^4 \Gamma \left(3 \gamma _1+1\right){}^2\right)^{-1}>0,
\end{eqnarray}
It cab further chracterized as 
\begin{equation}\label{18aa}
V''(\mathcal{X}_0)>0, \quad \Rightarrow \quad
\zeta(\mathcal{X}_0)\lambda_0^2-\eta_0>0.
\end{equation} 
Here, the coefficient of the EoS parameter is denoted with
$\zeta(\mathcal{X}_0)=\zeta_0$ and remaining terms are named as
$\mathcal{E}(\mathcal{X}_0)=\eta_0$.

The geometrical configuration of thin shell is explored by using
stable regions which can be expressed as\\\\
(i) For $\zeta_0<0$ $\Rightarrow$\quad
$\lambda_0^2<\eta_0/\zeta_0$:\\\\
(ii) For $\zeta_0>0$ $\Rightarrow$\quad $\lambda_0^2>\eta_0/\zeta_0$,\\\\
where 
\begin{eqnarray}\nonumber
\eta_0&=&h^4 \Gamma \left(3 \gamma _1+1\right){}^2 \left(\mathcal{X}_0 B(\mathcal{X}_0) \left(\mathcal{X}_0 \mathcal{Z}_0 ^2 B''(\mathcal{X}_0)+8 p_0 \mathcal{Z}_0  B'(\mathcal{X}_0)+4 \xi ^2 \left(8 \pi ^2 \mathcal{X}_0^2 \mathcal{Z}_0 ^4+6 p_0^2+3 p_0 \mathcal{Z}_0 \right)-12 p_0 (2 p_0+\mathcal{Z}_0 )\right)\right.\\\nonumber&+&\left.\mathcal{X}_0^2 \left(\mathcal{Z}_0  \left(\mathcal{X}_0 \mathcal{Z}_0  B''(\mathcal{X}_0) \left(\xi ^2 \left(16 \pi ^2 \mathcal{X}_0^2 \mathcal{Z}_0 ^2+1\right)-1\right)+B'(\mathcal{X}_0) \left(\mathcal{Z}_0  B'(\mathcal{X}_0)+2 \xi ^2 \left(-16 \pi ^2 \mathcal{X}_0^2 \mathcal{Z}_0 ^3+4 p_0+\mathcal{Z}_0 \right)\right.\right.\right.\right.\\\nonumber&-&\left.\left.\left.\left. 2 (4 p_0+\mathcal{Z}_0 )\right)\right)+4 p_0^2 \left(\xi ^4 \left(256 \pi ^4 \mathcal{X}_0^4 \mathcal{Z}_0 ^4+3\right)-6 \xi ^2+3\right)+2 p_0 \mathcal{Z}_0  \left(\xi ^4 \left(768 \pi ^4 \mathcal{X}_0^4 \mathcal{Z}_0 ^4+5\right)-10 \xi ^2+5\right)\right.\right.\\\nonumber&+&\left.\left.\mathcal{Z}_0 ^2 \left(\xi ^4 \left(768 \pi ^4 \mathcal{X}_0^4 \mathcal{Z}_0 ^4+1\right)-2 \xi ^2+1\right)\right)+2 p_0 B(\mathcal{X}_0)^2 (6 p_0+\mathcal{Z}_0 )\right)-2 h^2 \mathcal{X}_0^3 \mathcal{U} \xi ^2 \Gamma \left(3 \gamma _1+1\right) \\\nonumber&\times&\left(\mathcal{X}_0 \mathcal{Z}_0  \left(\mathcal{X}_0 \mathcal{Z}_0  B''(\mathcal{X}_0)+(8 p_0+6 \mathcal{Z}_0 ) B'(\mathcal{X}_0)\right)+B(\mathcal{X}_0) \left(24 p_0^2+28 p_0 \mathcal{Z}_0 +6 \mathcal{Z}_0 ^2\right)\right)+8 h^2 \mathcal{X}_0^4 \mathcal{U} \xi ^2 \Gamma \left(3 \gamma _1+1\right)\\\nonumber&\times& \left(3 (p_0+\mathcal{Z}_0 ) (2 p_0+\mathcal{Z}_0 )-\xi ^2 \left(-8 \pi ^2 \mathcal{X}_0^2 \mathcal{Z}_0 ^4+6 p_0^2+9 p_0 \mathcal{Z}_0 +3 \mathcal{Z}_0 ^2\right)\right)+4 \mathcal{X}_0^6 \mathcal{U}^2 \xi ^4 \left(12 p_0^2+26 p_0 \mathcal{Z}_0 +15 \mathcal{Z}_0 ^2\right),
\\\nonumber \zeta_0&=&2 \pi ^2 \mathcal{X}_0^6 \mathcal{Z}_0 ^4 \xi ^4 \left(\frac{(p_0+\mathcal{Z}_0 ) \left(B(\mathcal{X}_0)+\xi ^2 (\mathcal{X}_0-2 m)-\mathcal{X}_0\right)^2}{8 \pi ^2 \mathcal{X}_0^6 \mathcal{Z}_0 ^3 \xi ^4}-\frac{8 \pi  (p_0+\mathcal{Z}_0 ) \left(4 \pi  \mathcal{X}_0^2 \mathcal{Z}_0 \right)(\mathcal{X}_0)}{\mathcal{X}_0^2}\right).
\end{eqnarray}
Figures (\ref{p1})-(\ref{p4}) are very useful to explore the stable and unstable configurations of thin-shell around WH geometry by using Fuzzy black hole solution. The behavior of $\zeta_0$ represents the possibility of stable regions around the critical points. We are interested in examining the impact of dark fluid distribution on the stability of the shell by using linearized radial perturbation. It is found that the stability of the shell reduces by increasing the parameter; $\xi$ see Fig. (\ref{p1}). In Fig. (\ref{p2}), we observe the impact of $\gamma_1$ on the stability of the shell. It is found that the stable regions enhances before the position of critical points. For smaller values of $\beta$, we get the large regions stable regions Fig. (\ref{p3}). Stable regions reduces by increasing the value of $\beta$. For lower values of $\mathcal{Z}_0$, large stable regions have are found see Fig. (\ref{p4}).

\section{Comparative Analysis of Embedding Diagrams for Different Model Parameters}
\label{EmbeddingDiagrams}

\begin{figure}[h]
\centering
\subfloat[]{{\includegraphics[height=2.0 in, width=2.0 in]{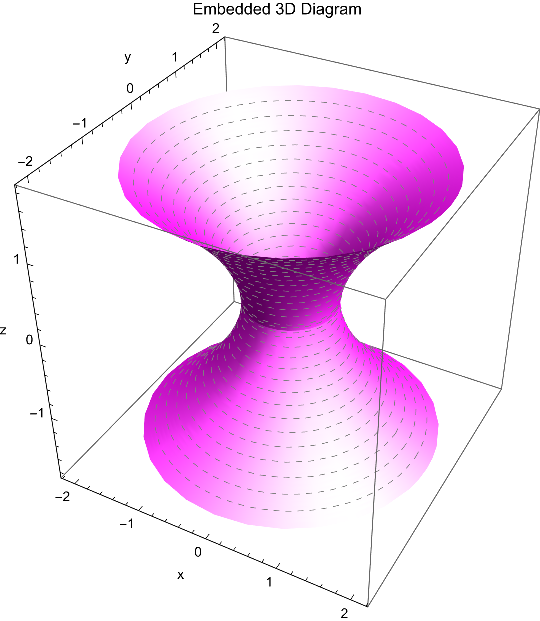}}}
\qquad
\subfloat[]{{\includegraphics[height=2.0 in, width=2.0 in]{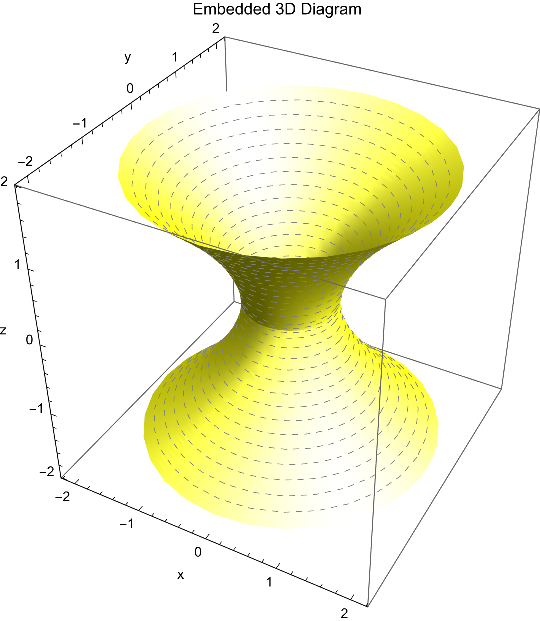}}}
\qquad
\subfloat[]{{\includegraphics[height=2.0 in, width=2.0 in]{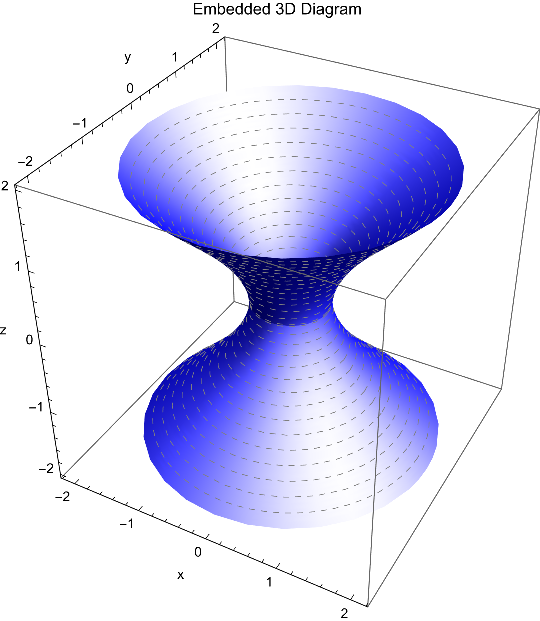}}}
\qquad
\subfloat[]{{\includegraphics[height=3.4 in, width=3.4 in]{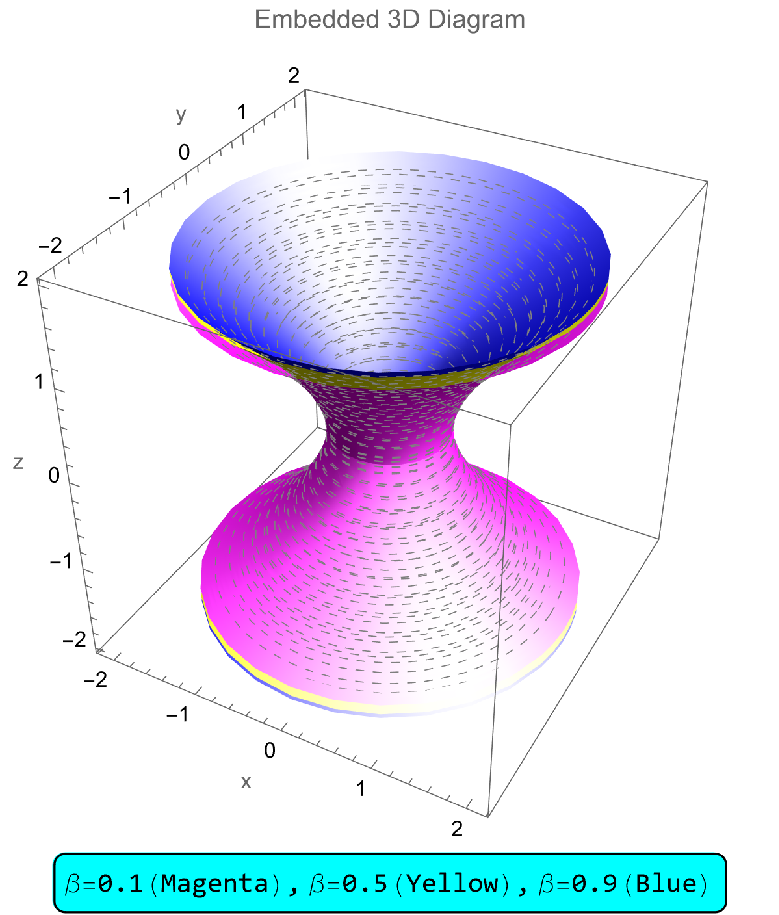}}}
\caption{Three dimensional embedding diagrams of the Einasto
dark matter supported fuzzy WH geometry for parameter sets listed in
Table~\ref{tab:embedding_parameters}, with fixed $\gamma_{1}=3.33$ and
$a_{0}=0.6$, while panels (a), (b), and (c) correspond to
$(\rho_{0},\beta,\xi)=(0.1,0.1,0.51)$, $(0.3,0.5,0.71)$, and $(0.5,0.9,0.91)$, respectively, whereas panel (d)
provides a combined representation of the three embedding surfaces for
direct comparison. The upper as well as lower branches connect smoothly at WH throat and illustrate the dependence of embedding depth and radial
deformation on model parameters.}
\label{fig:embedding}
\end{figure}

In this section, we investigate intrinsic spatial geometry of constructed fuzzy WH configuration through embedding diagrams, while embedding formalism provides a useful geometrical visualization of curvature of WH spacetime and, in particular, illustrates how throat,
flare-out behavior, embedding depth, as well as radial extension are influenced by parameters of the Einasto dark matter distribution and global monopole charge. Since, present WH configuration is static and
spherically symmetric, its spatial geometry can conveniently be examined on a constant time hypersurface, $t=\mathrm{constant}$, while the corresponding spatial line element obtained from WH metric can be expressed as
\begin{equation}
dS^{2}=\xi^{-2}\left(1-\frac{B(r)}{r}\right)^{-1}dr^{2}+r^{2}\left(d\theta^{2}+\sin^{2}\theta\,d\phi^{2}\right),
\label{eq:embedding_spatial_metric}
\end{equation}
where $B(r)$ denotes the Einasto profile-based shape function and $\xi$
represents global monopole parameter, whereas throat radius $a_{0}$ is
determined through standard condition $B(a_{0})=a_{0}$. To visualize the WH geometry, we restrict our attention to the equatorial plane $\theta=\pi/2$, while under this restriction, Eq.~(\ref{eq:embedding_spatial_metric})
reduces to
\begin{equation}
dS^{2}=\xi^{-2}\left(1-\frac{B(r)}{r}\right)^{-1}dr^{2}+r^{2}d\phi^{2}.
\label{eq:equatorial_embedding_metric}
\end{equation}
The resulting two dimensional curved geometry can be embedded in a
three dimensional Euclidean space described by cylindrical coordinates
$(r,\phi,z)$, while these coordinates are related to corresponding Cartesian coordinates through
$x=r\cos\phi,~y=r\sin\phi,~z=z(r).$ The Euclidean line element is given by
\begin{equation}
dS^{2}=dr^{2}+r^{2}d\phi^{2}+dz^{2}.
\label{eq:euclidean_metric}
\end{equation}
Since the embedding surface is characterized by $z=z(r)$, one can write
$dz=(dz/dr)dr$, and hence above line element takes form
\begin{equation}
dS^{2}=\left[1+\left(\frac{dz}{dr}\right)^{2}\right]dr^{2}+r^{2}d\phi^{2}.
\label{eq:induced_embedding_metric}
\end{equation}
Comparing Eqs.~(\ref{eq:equatorial_embedding_metric}) and
(\ref{eq:induced_embedding_metric}), we obtain the differential equation
governing the embedding profile,
\begin{equation}
\frac{dz}{dr}=\pm\left[\frac{1}{\xi^{2}\left(1-B(r)/r\right)}-1\right]^{1/2}.
\label{eq:embedding_derivative}
\end{equation}
The positive as well as negative signs correspond to the upper and lower branches of embedded WH surface, respectively, while the embedding function can therefore be expressed in integral form as
\begin{equation}
z(r)=\pm\int_{a_{0}}^{r}\left[\frac{1}{\xi^{2}\left(1-B(\tilde r)/\tilde r\right)}-1\right]^{1/2}
d\tilde r.
\label{eq:embedding_integral}
\end{equation}
For numerical evaluation, the integration is initiated slightly outside the throat, i.e., $z(a_{0}+\epsilon)=0$, $\epsilon>0$, where $\epsilon$ denotes a sufficiently small positive quantity introduced to avoid coordinate divergence occurring exactly at throat. An important geometrical feature of WH throat is vertical tangent of embedding profile, since $B(a_{0})=a_{0}$, the denominator appearing in
Eq.~(\ref{eq:embedding_derivative}) approaches zero as $r\rightarrow a_{0}$, and consequently $|dz/dr|$ becomes very large. This behavior geometrically represents narrow connecting region between two branches of WH and is closely associated with flare-out property, however away from throat, evolution of $z(r)$ provides a direct measure of how rapidly the spatial geometry opens and approaches its exterior behavior. To investigate sensitivity of embedding geometry to the Einasto
dark matter and global monopole parameters, we consider three representative parameter combinations, while the Einasto index and throat radius are fixed at
$\gamma_{1}=3.33$ and $a_{0}=0.6$, respectively, whereas $\rho_{0}$, $\beta$, and $\xi$ are varied. The parameter sets employed in constructing embedding diagrams are summarized in Table~\ref{tab:embedding_parameters}.
\begin{table}[H]
\centering
\begin{tabular}{|c|c|c|c|c|c|c|}
\hline
\hline
~~~\textbf{Case}~~~~ & ~~~$\gamma_{1}$~~~ & ~~~$\rho_{0}$~~~ & ~~~$a_{0}$~~~ &~~~ $\beta$~~~ &~~ ~$\xi$~~~ & ~~~Color ~(Plot)~~~ \\
\hline
\hline
Fig. \ref{fig:combined3D} (a) & $3.33$ & $0.1$ & $0.6$ & $0.1$ & $0.51$ & Magenta \\
\hline
Fig. \ref{fig:combined3D} (b) & $3.33$ & $0.3$ & $0.6$ & $0.5$ & $0.71$ & Yellow \\
\hline
Fig. \ref{fig:combined3D} (c) & $3.33$ & $0.5$ & $0.6$ & $0.9$ & $0.91$ &  Blue \\
\hline
\hline
\end{tabular}
\caption{Parameter sets employed for constructing the embedding diagrams of the Einasto dark matter supported fuzzy WH configuration.}\label{tab:embedding_parameters}
\end{table}
The three dimensional embedding surfaces corresponding to the parameter
sets listed in Table~\ref{tab:embedding_parameters} are displayed in
Fig.~\ref{fig:embedding}. Panels (a)-(c) present individual embedding geometries, whereas panel (d) provides a combined representation that facilitates a direct comparison among three configurations, while in every case, the upper and lower branches are smoothly connected through throat at $r=a_{0}$, thereby providing a clear geometrical representation of the two sided WH structure. For the parameter set displayed in Fig.~\ref{fig:embedding}(a), namely $(\rho_{0},\beta,\xi)=(0.1,0.1,0.51)$, the embedding surface exhibits a
comparatively shallow profile with moderate radial deformation, however, the smaller central density and scale parameter correspond to a less pronounced curvature of embedded surface within the parameter range considered. The throat remains well defined and joins the two branches smoothly. For Fig.~\ref{fig:embedding}(b), corresponding to
$(\rho_{0},\beta,\xi)=(0.3,0.5,0.71)$, the embedding geometry becomes more pronounced compared with that shown in panel (a), while the surface develops a larger embedding depth and a broader radial extension, indicating that combined variation of matter distribution parameters as well as monopole parameter appreciably modifies intrinsic spatial curvature of the WH. The third configuration, displayed in Fig.~\ref{fig:embedding}(c), corresponds to
$(\rho_{0},\beta,\xi)=(0.5,0.9,0.91)$, whereas for this parameter choice, the embedding surface exhibits most extended geometrical deformation among three cases considered. The upper and lower branches remain smoothly connected at throat, while profile becomes wider and more strongly curved over displayed radial region, while this behavior demonstrates sensitivity of WH spatial geometry to simultaneous variations of $\rho_{0}$, $\beta$, and $\xi$. Panel (d) of Fig.~\ref{fig:embedding} displays the three embedding geometries in a common three dimensional representation. This comparison clearly illustrates that changing the Einasto density parameters together with the global monopole charge modifies embedding depth, radial extension, and overall curvature of WH geometry, in particular, the magenta, yellow, and blue surfaces correspond, respectively, to
$(\rho_{0},\xi,\beta)=(0.1,0.51,0.1),$ $(0.3,0.71,0.5)$, $(0.5,0.91,0.9)$, while $\gamma_{1}=3.33$ and $a_{0}=0.6$ are kept fixed. The embedding analysis therefore confirms that Einasto profile parameters and global monopole contribution play an important role in determining the intrinsic spatial geometry of constructed fuzzy WH, although all  considered configurations preserve a smooth throat joining upper and lower sheets, their embedding depth and radial deformation vary considerably with adopted parameter sets. Hence, Fig.~\ref{fig:embedding} provides a direct geometrical visualization of how modifications in dark matter distribution and topological charge reshape WH geometry while
maintaining the characteristic throat structure.

\section{Timelike geodesics and Trajectories}\label{geometry}

To determine geodesic equations on the equatorial plane where \( \theta=\frac{\pi}{2} \), we analyze the Lagrangian equation of motion in WH geometry given as \( \mathcal{L} = \frac{1}{2} g_{\alpha \beta} \frac{dx^\alpha}{d\tau} \frac{dx^\beta}{d\tau} \). Thus, first order geodesics are derived from the so-called canonical conserved momenta and are expressed as
\begin{eqnarray}
    \dot{t} &=& E e^{-2 \Phi(r)},
    \label{geo06} \\
    \dot{\phi} &=& L/r^2,
    \label{geo07} \\
    \dot{r}^2 &=& \xi ^2\left( 1-\frac{B(r)}{r} \right)\left(E^2 e^{-2 \Phi(r)} -\frac{L^2}{r^2}+\varepsilon \right),
    \label{geo08}
\end{eqnarray}
In this context, $\varepsilon=1, 0, -1$ represents the geodesics that are spacelike, null, and timelike, respectively, while $E$ and $L$ stand for the energy and angular momentum of the particle, respectively. The overdot indicates differentiation with respect to the affine parameter for the null geodesics and with respect to the proper time for the geodesics that are timelike. Concerning radial null geodesics, Equation \eqref{geo08} simplifies to
\begin{eqnarray}
    \dot{r}^2 = \xi ^2\left( 1-\frac{B(r)}{r} \right)\left(E^2 e^{-2 \Phi(r)} -\frac{L^2}{r^2}+\varepsilon \right).
    \label{geo09}
\end{eqnarray}

One may begin to discuss the photon orbits by introducing,
\begin{eqnarray}
    \left( \frac{dr}{d\phi} \right)^2 = \xi ^2\left( 1-\frac{B(r)}{r} \right)\left( \frac{E^2}{L^2}r^4 e^{-2\Phi(r)}-{r^2}+\varepsilon r^4 \right).
    \label{geo10}
\end{eqnarray}

Assuming that the spacetime geometry is generated by a point mass source of radius $(r_s)$, photons coming from spatial infinity will not reach the surface as long as $(r_0 > r_s)$ is satisfied. Here, the radial speed is zero; that is, $( \dot{r}^2 = 0 )$. The quantity $(r_0)$ is the turning point, or the minimum distance of the approach of the photon path \cite{mishra2018}
\begin{eqnarray}
    \frac{L^2}{E^2}=r_0^2 e^{-2\Phi(r_0)}+\varepsilon r_0^2~~~~~~(\text{if}~\frac{B(r)}{r}\ne 1~\text{for~any}~r>r_s).
    \label{geo11}
\end{eqnarray}
Considering first some basic forms of movement of massive particles across the spacetime of a WH, for ultrastatic WHs, the starting velocity, location, and acceleration of radial timelike geodesics were studied by Cataldo et al. \cite{cataldo2017}. Extending this study in the radial direction, the complete non radial timelike geodesic acceleration can be obtained from the first-order radial timelike geodesic equation in \eqref{geo08}. We get
\begin{equation}
    \ddot{r}= \frac{1}{2r^4}\left(\xi ^2 \left(r B'(r) \left(L^2-r^2 \left(E^2+\varepsilon \right)\right)-3 L^2 B(r)+r^2 B(r) \left(E^2+\varepsilon \right)+2 L^2 r\right)\right) .
    \label{geo14}
\end{equation}

At this point, we are about to study the circular timelike geodesics which has some particular point of interest. For usual cases, it is well established that circular timelike geodesics can only exist at the throat and for them radial velocity vanishes, that means $ \dot{r}=0 $. 
As at $ r=b_0 $, the acceleration vanishes; one can always verify that circular timelike geodesics can only exist at the throat. For ultrastatic WHs, the radial geodesic equation in terms of proper distance can be written as
\begin{equation}
    \left( \frac{dl}{ds} \right)^2 = E^2-V(L,l) ,
    \label{geo16}
\end{equation}
in terms of  proper distance and  conserved momentum, the potential function is given as 
\begin{equation}
    V(L,l)= \frac{L^2}{r(l)^2}+\varepsilon ,
    \label{geo17}
\end{equation}
For $L \ne 0$, this expression remains strictly positive while tending to zero as $l \rightarrow  \pm \infty$. Here, we get
\begin{equation}
    \frac{dV}{dl}\Big|_{l=0}=0 ,
    \label{geo18}
\end{equation}
In eq. \eqref{geo18}, the potential has a global maximum at the throat. If $ r(l) $ is concave up, the only turning point occurs at $ l=0 $, which separates the bound orbits from the unbounded orbits that transmit. We have some stable circular geodesic orbits if $ r(l) $ is not necessarily concave up, and the potential well exhibits oscillatory behavior around turning points \cite{taylor2014, sarbach2012}.

\begin{figure}[h]\centering
\subfloat[L=15,E=8]{{\includegraphics[height=2.0 in, width=2.0 in]{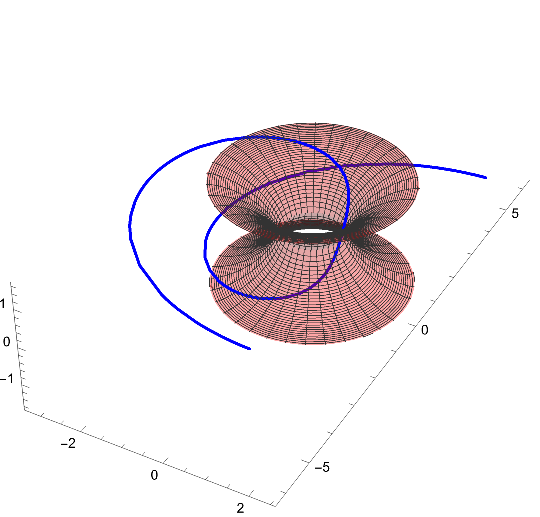}}}
\qquad
\subfloat[L=25,E=8]{{\includegraphics[height=2.0 in, width=2.0 in]{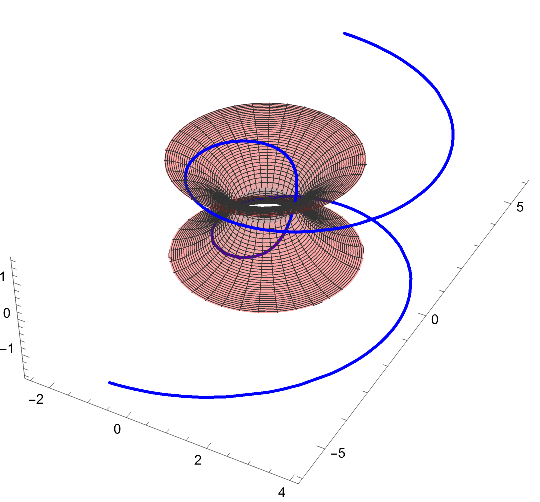}}}
\qquad
\subfloat[L=35,E=8]{{\includegraphics[height=2.0 in, width=2.0 in]{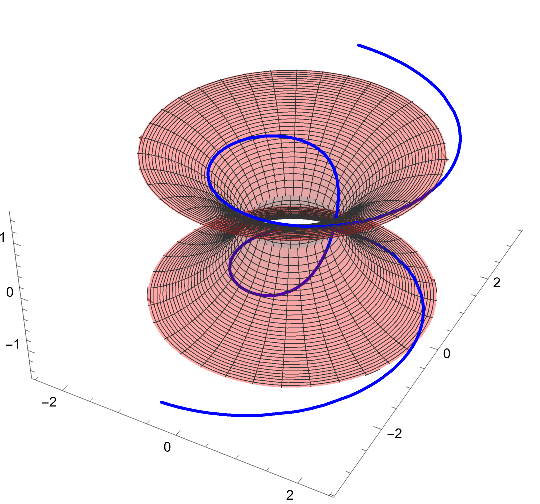}}}
\caption{Escape orbits for $\xi=0.7$ with different values of $L$.} 
\label{fig:combined3D}
\centering
\subfloat[L=15,E=10]{{\includegraphics[height=2.0 in, width=2.0 in]{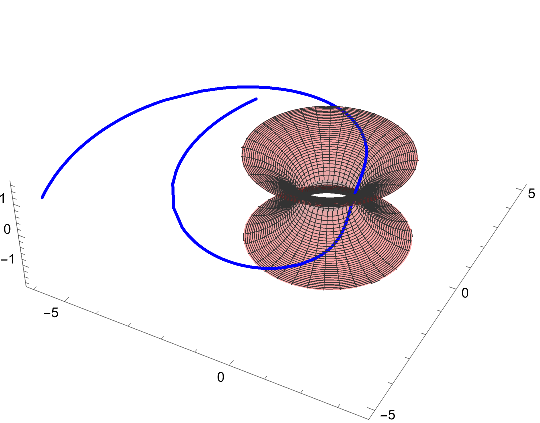}}}
\qquad
\subfloat[L=15,E=15]{{\includegraphics[height=2.0 in, width=2.0 in]{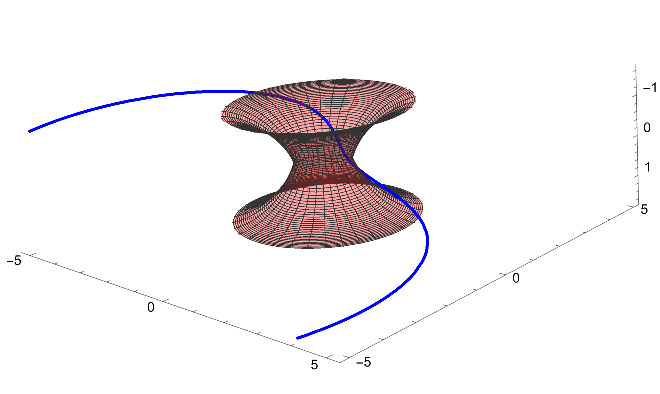}}}
\qquad
\subfloat[L=15,E=20]{{\includegraphics[height=2.0 in, width=2.0 in]{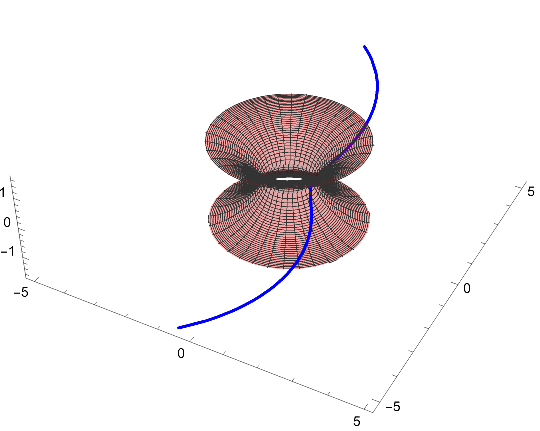}}}
\caption{Escape orbits for $\xi=0.7$ with different values of $E$.}
\end{figure}

\begin{figure}[h]
\centering
\subfloat[L=15,E=8]{{\includegraphics[height=2.0 in, width=2.0 in]{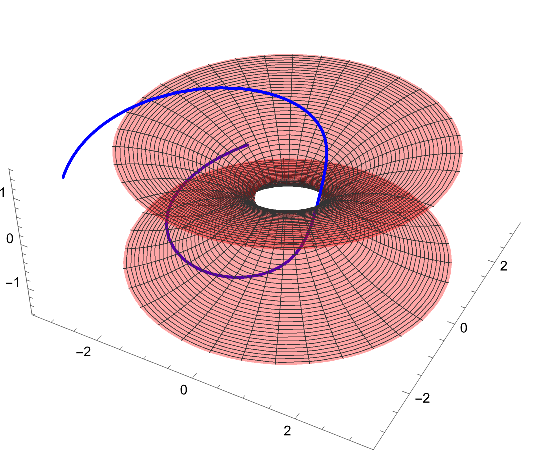}}}
\qquad
\subfloat[L=25,E=8]{{\includegraphics[height=2.0 in, width=2.0 in]{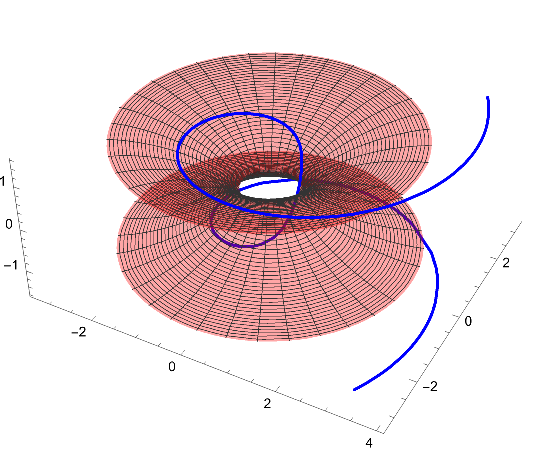}}}
\qquad
\subfloat[L=35,E=8]{{\includegraphics[height=2.0 in, width=2.0 in]{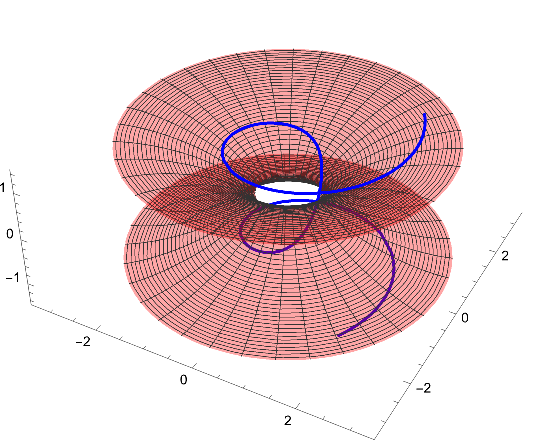}}}
\caption{Escape orbits for $\xi=0.8$ with different values of $L$.} 
\label{fig:combined3D}
\centering
\subfloat[L=15,E=10]{{\includegraphics[height=2.0 in, width=2.0 in]{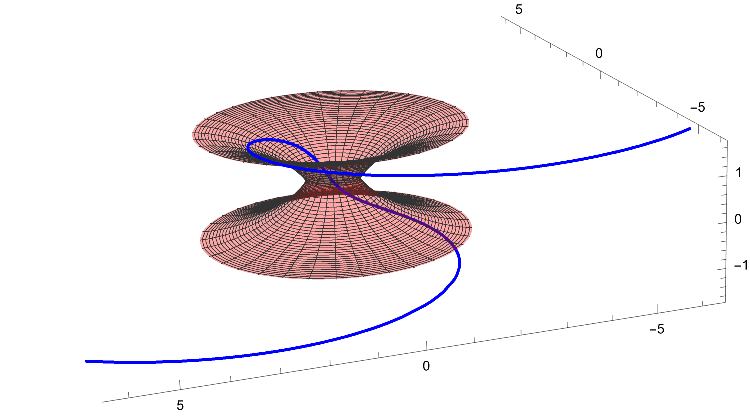}}}
\qquad
\subfloat[L=15,E=15]{{\includegraphics[height=2.0 in, width=2.0 in]{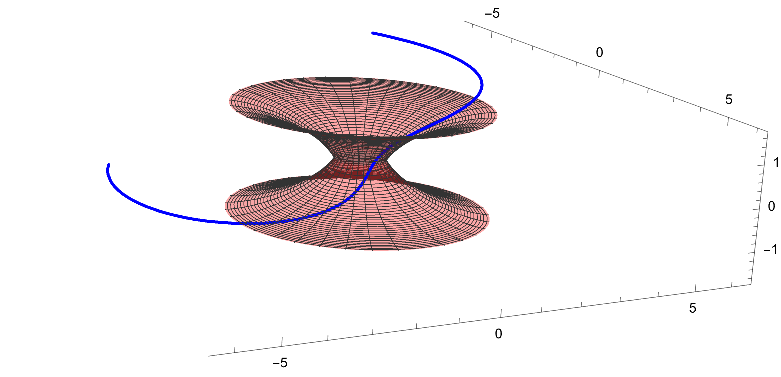}}}
\qquad
\subfloat[L=15,E=20]{{\includegraphics[height=2.0 in, width=2.0 in]{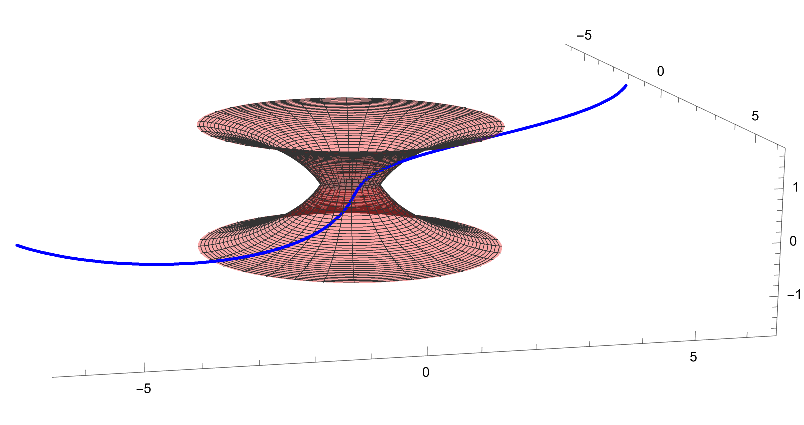}}}
\caption{Escape orbits for $\xi=0.8$ with different values of $E$.}
\end{figure}

Geodesic paths can be categorized into bounded and unbounded orbits, however, both types of trajectories are classified as timelike geodesics. Bound orbits are geodesics that originate from spatial infinity, arrive at the throat, and then reflect back out to spatial infinity. Unbounded trajectories, however, are able to cross the throat to other universes and thus other spatial infinities. When considering traversable WHs, it is critically important that the trajectory of the massive (timelike) particles from one universe has to cross the throat to the other universe (or to some distant point in the same universe). This type of unbounded trajectory is called an escape orbit. Recalling Eq. \eqref{geo08} in terms of proper distance, we have
\begin{equation}
    \dot{l}^2 = \left(E^2 e^{-2\Phi(r)}-\frac{L^2}{r^2}-1 \right) .
    \label{geo20}
\end{equation}
The bound orbit condition seems to be: deflecting particles coming from infinity after they get close to the throat at the minimum distance \( b_0 \):
\begin{equation}
    \left(E^2 e^{-2\Phi(b_0)}-\frac{L^2}{b_0^2}-1 \right) \le 0,
    \label{geo21}
\end{equation}
which is simplified to,
\begin{equation}
    \frac{L}{\sqrt{E^2 e^{-2\Phi(b_0)}-1}} \ge b_0.
    \label{geo22}
\end{equation}

Figures 12-15 illustrate the escape trajectories of test particles in the WH spacetime for different values of the angular momentum $L$, energy $E$, and monopole charge parameter $\xi$. Here, $\xi$ represents the monopole charge characterizing the WH geometry, which directly modifies the curvature of the spacetime and consequently the structure of the effective potential. From Figs.~12 and 14, where the energy is fixed and $L$ varies, it is evident that increasing the angular momentum enhances the centrifugal barrier term $L^{2}/r^{2}$. As a result, the particle experiences stronger azimuthal motion and is prevented from penetrating deeply into the gravitational well near the WH throat. For smaller $L$, the particle approaches closer to the throat before escaping, producing tighter spiral-like trajectories. In contrast, larger $L$ values lead to wider deflection and earlier escape at larger radial distances. The influence of the monopole charge $\xi$ is manifested through the deformation of the potential well; increasing $\xi$ modifies the curvature around the throat and shifts the escape radius, indicating that the monopole charge directly affects the strength of the gravitational attraction. Figures 13 and 15 demonstrate the effect of varying the particle energy while keeping $L$ fixed. For lower energies, the particle remains longer in the vicinity of the effective potential well, performing multiple windings before escaping. As the energy increases, the particle more readily overcomes the potential barrier, leading to faster escape with fewer revolutions. At sufficiently high energy, the motion resembles scattering-type trajectories with minimal orbital winding. The monopole charge $\xi$ alters the depth and width of the effective potential, thereby modifying the critical energy required for escape and the overall deflection angle.  Overall, the escape behavior results from the competition between gravitational attraction determined by the WH geometry (through $\xi$) and the centrifugal repulsion governed by $L$, while the total energy $E$ controls the transition from quasi-bound motion to unbound scattering trajectories.

\section{Mass-Radius Relation and Less Complex Nature of Fuzzy Wormhole Models}\label{MassRadius_ComplexityFactor}

Let us present a unified analysis of an active gravitational mass and complexity nature by using $Y_{TF}$ factor associated with fuzzy WH configurations. These two aforementioned physical quantities complement each other in understanding how matter distribution, anisotropic stresses, and energy density variations collectively govern the internal structural behavior of the system. However, together these quantities provide a comparative picture of how the fuzzy WH evolves from its throat to the outer regions. The active gravitational mass $M_a$ quantifies the amount of matter enclosed between the WH throat radius $a_0$ and any radial position $r$, whereas for a given fuzzy WH density profile, it is defined as
\begin{align}\nonumber
M_{a}&=4\pi\int_{a_0}^{r}\rho(r)\, r^{2}\,dr=4\pi\int_{a_0}^{r} \frac{1}{r^2}\left[-\xi ^2+\xi ^2 \left(-\frac{1}{\xi ^2}-\xi ^{-2}\rho_0 r^2 \sinh \left(\left(\frac{r}{\beta}\right)^{1/\gamma_{1}}\right)\right.\right.
\\\label{a27}&\left.\left.+\xi ^{-2}\rho_0 r^2 \cosh \left(\left(\frac{r}{\beta}\right)^{1/\gamma_{1}}\right)+1\right)+1\right] r^{2}dr.
\end{align}
In contrast to the mass profile, the complexity factor offers insight into the internal structural organization of the matter content, while a physical system is considered complex when multiple interacting components such as anisotropic stresses, energy density gradients, and pressure variations prevent a simple structural configuration. Herrera \cite{herrera2018new} introduced the notion of a complexity factor for spherically symmetric, static fluids in EGR, showing that a configuration is minimally complex (i.e., $Y_{TF}=0$) when the fluid possesses both homogeneous energy density and isotropic pressure. Interestingly, even when energy density is inhomogeneous and pressure stresses are anisotropic, a vanishing complexity factor may still arise if the contributions of these two effects cancel each other out \cite{andrade2023anisotropic,al2023complexity,naseer2024implications}. For the fuzzy WH model, the complexity factor is defined as
\begin{align}\nonumber
Y_{TF}&=P_{\text{rad}}-P_{\text{tan}}-\frac{1}{2r^{3}}\int_{a_0}^{r} r^{3}\rho'(r)\,dr
=P_{\text{rad}}-P_{\text{tan}}-\frac{1}{2r^{3}}\int_{a_0}^{r} \left[ \frac{1}{r^2}\left\{-\xi ^2+\xi ^2 \left(-\frac{1}{\xi ^2}-\xi ^{-2}\rho_0 r^2 \sinh \left(\left(\frac{r}{\beta}\right)^{1/\gamma_{1}}\right)\right.\right.\right.
\\\label{a35}&\left.\left.\left.+\xi ^{-2}\rho_0 r^2 \cosh \left(\left(\frac{r}{\beta}\right)^{1/\gamma_{1}}\right)+1\right)+1\right\}\right]'  r^{3} dr
\end{align}
which directly encodes the effects of pressure anisotropy and density inhomogeneity. By adopting a constant redshift function and inserting the expressions for $\rho$, $P_{\text{rad}}$, $P_{\text{tan}}$, and the shape function, we obtain
\begin{align}\label{a28}
M_{a}=4 \pi  \rho_0 \left(\gamma_{1} a_{0}^3 \left(\left(\frac{a_{0}}{\beta}\right){}^{1/\gamma_{1}}\right){}^{-3 \gamma_{1}} \Gamma \left(3 \gamma_{1},\left(\frac{a_{0}}{\beta}\right){}^{1/\gamma_{1}}\right)-\gamma_{1} r^3 \left(\left(\frac{r}{\beta}\right)^{1/\gamma_{1}}\right)^{-3 \gamma_{1}} \Gamma \left(3 \gamma_{1},\left(\frac{r}{\beta}\right)^{1/\gamma_{1}}\right)\right).
\end{align}
and
\begin{align}\nonumber
&Y_{TF} =\frac{1}{2} \left(\rho_0 \left(-\frac{a_{0}^3 \left(\left(\frac{a_{0}}{\beta}\right){}^{1/\gamma_{1}}\right){}^{-3 \gamma_{1}} \left(3 \gamma_{1} \Gamma \left(3 \gamma_{1},\left(\frac{a_{0}}{\beta}\right){}^{1/\gamma_{1}}\right)-\Gamma \left(3 \gamma_{1}+1,\left(\frac{a_{0}}{\beta}\right){}^{1/\gamma_{1}}\right)\right)}{r^3}+e^{-\left(\frac{r}{\beta}\right)^{1/\gamma_{1}}}+3 \gamma_{1} \right.\right.
\\\label{a36}&\times\left.\left.\left(\left(\frac{r}{\beta}\right)^{1/\gamma_{1}}\right)^{-3 \gamma_{1}}\Gamma \left(3 \gamma_{1},\left(\frac{r}{\beta}\right)^{1/\gamma_{1}}\right)-\left(\left(\frac{r}{\beta}\right)^{1/\gamma_{1}}\right)^{-3 \gamma_{1}} \Gamma \left(3 \gamma_{1}+1,\left(\frac{r}{\beta}\right)^{1/\gamma_{1}}\right)\right)-\frac{3 a_{0}}{r^3}\right).
\end{align}

\begin{figure}[H]
\centering
\subfloat[]{{\includegraphics[height=2.8in,width=3.4in]{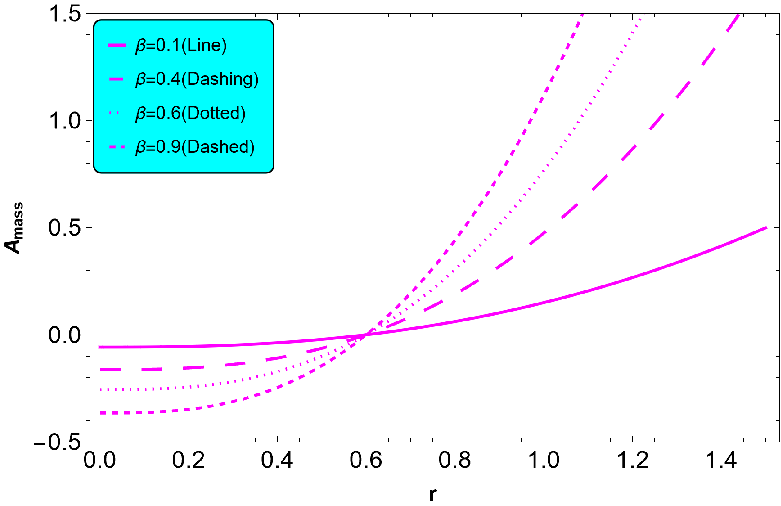}}}
\qquad
\subfloat[]{{\includegraphics[height=2.8in,width=3.4in]{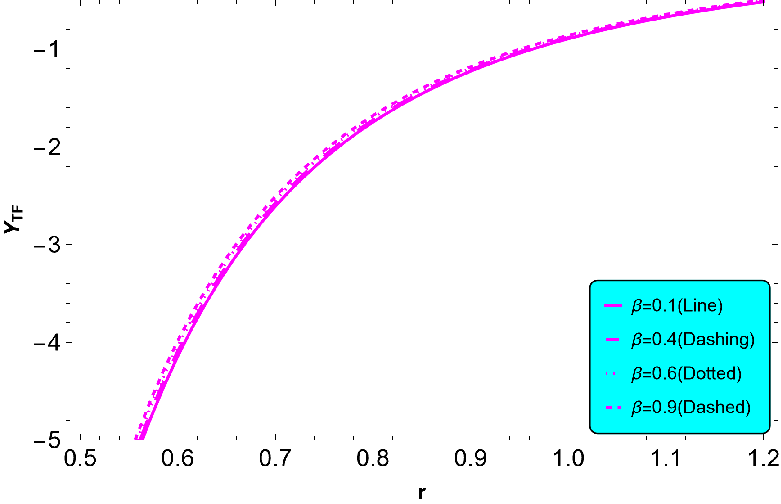}}}
\caption{For panel (a) active gravitational mass $M_{a}$ for fuzzy WHs with $a_0=0.6$, shown for the Einasto index $\gamma_{1}=3.33$ and panel (b) complexity factor $Y_{TF}$ versus radial coordinate $r$ for the same configurations.}
\label{massComplexity}
\end{figure}
From the schematic representation one can observe that: 
\begin{description}
  \item[a)] Fig.~\ref{massComplexity}(a) displays the radial evolution of $M_a$ for different values of the Einasto index $\gamma_{1}$ and an increase in $\gamma_{1}$ leads to a stronger rise in $M_a$, demonstrating that the gravitational mass grows more rapidly as the density distribution becomes sharper.
  \item[b)] A narrow region near the throat shows slightly negative values of $M_a$, signalling the presence of exotic matter required to sustain the traversable WH geometry; however, as $r$ increases, the WH mass becomes positive and monotonically increasing. 
  \item[c)] One can observe from Figs.~\ref{massComplexity}(a)-(b) which reveal an interesting structural interplay as an active gravitational mass increases steadily with respect to radial coordinate $r$, the complexity factor decreases and approaches zero away from the throat.
  \item[d)] This complementary behavior reflects how matter becomes progressively better organized at large $r$, with pressure anisotropy diminishing and density gradients weakening.
\end{description}
Consequently, the complexity is highest near the WH throat where exotic matter, strong anisotropy, and rapid density variations coexist but it gradually fades at larger radii, ultimately indicating an effective transition toward isotropy and structural simplicity. This comparison highlights that mass accumulation and structural complexity evolve in opposite manners in fuzzy WH geometries.

\begin{figure}[H]
\centering
\subfloat[]{{\includegraphics[height=2.8 in, width=3.4 in]{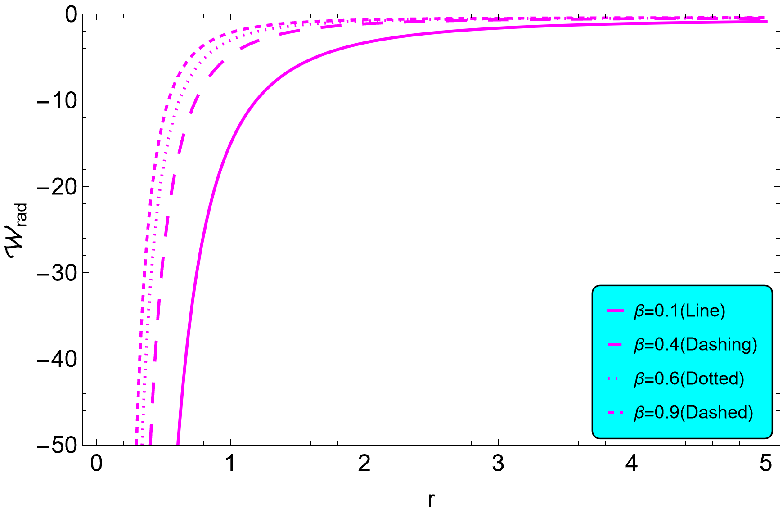}}}
\qquad
\subfloat[]{{\includegraphics[height=2.8 in, width=3.4 in]{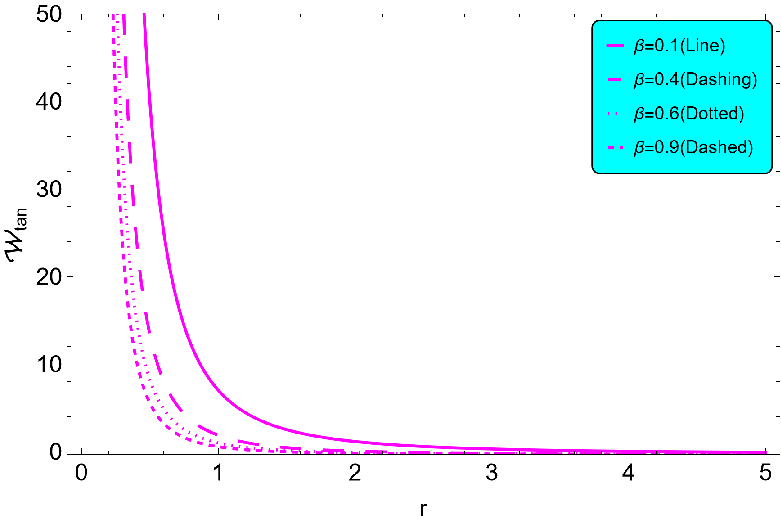}}}
\caption{Plots in panel (a) radial component of EoS vs $r$, and panel (b) tangential component vs $r$.}
\label{EoS}
\end{figure}

The EoS for the radial and tangential pressure components are defined as
\[
\mathcal{W}_{rad}=\frac{P_{\text{rad}}}{\rho}, \qquad  
\mathcal{W}_{tan}=\frac{P_{\text{tan}}}{\rho},
\]
which describe the ratio between each pressure component and the corresponding energy density. Using the Einasto-type density distribution adopted in this model, the explicit expressions for the radial and tangential EoS parameters are derived as
\begin{align}\nonumber
\mathcal{W}_{rad}&=-\frac{e^{\left(\frac{r}{\beta}\right)^{1/\gamma_{1}}}}{\rho_0 r^3} \left(\xi ^2 \left(\frac{a_{0} \left(\left(\frac{a_{0}}{\beta}\right){}^{1/\gamma_{1}}\right){}^{-3 \gamma_{1}} \left(\gamma_{1} \rho_0 a_{0}^2 \Gamma \left(3 \gamma_{1},\left(\frac{a_{0}}{\beta}\right){}^{1/\gamma_{1}}\right)+\left(\left(\frac{a_{0}}{\beta}\right){}^{1/\gamma_{1}}\right){}^{3 \gamma_{1}}\right)}{\xi ^2}+\gamma_{1} \rho_0 \left(-r^3\right) \right.\right.
\\\label{a25}&\times\left.\left.\left(\left(\frac{r}{\beta}\right)^{1/\gamma_{1}}\right)^{-3 \gamma_{1}}\frac{ \Gamma \left(3 \gamma_{1},\left(\frac{r}{\beta}\right)^{1/\gamma_{1}}\right)+\xi ^2 r-r}{\xi ^2}\right)+\xi ^2 (-r)+r\right)
,
\\\nonumber
\mathcal{W}_{tan}&=\frac{\xi ^2 e^{\left(\frac{r}{\beta}\right)^{1/\gamma_{1}}}}{2 \rho_0 r^3} \left(\frac{a_{0} \left(\left(\frac{a_{0}}{\beta}\right){}^{1/\gamma_{1}}\right){}^{-3 \gamma_{1}} \left(\gamma_{1} \rho_0 a_{0}^2 \Gamma \left(3 \gamma_{1},\left(\frac{a_{0}}{\beta}\right){}^{1/\gamma_{1}}\right)+\left(\left(\frac{a_{0}}{\beta}\right){}^{1/\gamma_{1}}\right){}^{3 \gamma_{1}}\right)}{\xi ^2}+\gamma_{1} \rho_0 \left(-r^3\right) \right.
\\\label{a26}&\times\left.\left.\left(\left(\frac{r}{\beta}\right)^{1/\gamma_{1}}\right)^{-3 \gamma_{1}}\frac{ \Gamma \left(3 \gamma_{1},\left(\frac{r}{\beta}\right)^{1/\gamma_{1}}\right)+\xi ^2 r-r}{\xi ^2}\right)-\frac{r \left(\xi ^2+\rho_0 r^2 e^{-\left(\frac{r}{\beta}\right)^{1/\gamma_{1}}}-1\right)}{\xi ^2}\right).
\end{align}
The NEC can be expressed conveniently in terms of the radial as well as tangential equation of state parameters as
\begin{equation}
\rho+P_{\rm rad}=\rho\left(1+W_{\rm rad}\right),\quad
\rho+P_{\rm tan}=\rho\left(1+W_{\rm tan}\right).
\end{equation}
For the configurations considered in present analysis, energy density remains positive in region of interest, therefore, sign of
NEC combinations is directly determined by quantities $1+W_{\rm rad}$ as well as $1+W_{\rm tan}$, in particular, radial NEC is
violated whenever
\begin{equation}
1+W_{\rm rad}<0,\quad \text{or equivalently} \quad W_{\rm rad}<-1,
\end{equation}
whereas tangential NEC is violated if
\begin{equation}
1+W_{\rm tan}<0,\quad \text{or equivalently} \quad W_{\rm tan}<-1.
\end{equation}
On the other hand, the inequalities
\begin{equation}
1+W_{\rm rad}\geq 0, \quad 1+W_{\rm tan}\geq 0,
\end{equation}
correspond to the satisfaction of the radial and tangential NECs,
respectively, while the graphical behavior of EoS parameters is presented in Fig.~\ref{EoS}, whereas panel~(a) shows that $W_{\rm rad}$ falls below phantom divide $W_{\rm rad}=-1$ within a finite radial interval close to the WH throat. Consequently,
\begin{equation}
\rho+P_{\rm rad}=\rho(1+W_{\rm rad})<0,
\end{equation}
which confirms violation of radial NEC in this region, such a violation is a characteristic feature of traversable WH geometries and indicates presence of exotic stresses required to support throat. In contrast, panel~(b) shows that $W_{\rm tan}$ remains predominantly
greater than $-1$ for the adopted parameter choices, hence,
\begin{equation}
\rho+P_{\rm tan}=\rho(1+W_{\rm tan})>0,
\end{equation}
and the tangential NEC remains satisfied over the corresponding radial
domain, in particular, positive values of $W_{\rm tan}$ displayed in
Fig.~\ref{EoS}(b) imply $1+W_{\rm tan}>0$, which is fully consistent with satisfaction of NEC in tangential direction. The combined behavior of $W_{\rm rad}$ as well as $W_{\rm tan}$ therefore reveals an anisotropic matter distribution in which exotic character is primarily associated with the radial pressure component. The radial EoS crosses phantom divide near throat and produces the required NEC violation, whereas tangential sector can retain the standard energy condition for considered parameter ranges. This result is consistent with the earlier energy condition analysis, where the exotic contribution is concentrated mainly in the vicinity of WH throat and gradually weakens with increasing radial distance.
From Fig.~\ref{EoS}, the following features can be observed:
\begin{itemize}
\item[(a)] The radial EoS parameter $W_{\rm rad}$ exhibits values below
$-1$ in a finite neighborhood of WH throat, while this corresponds to the
phantom regime and confirms violation of radial NEC, $\rho+P_{\rm rad}<0$.
\item[(b)] As radial coordinate increases, $W_{\rm rad}$ approaches value $-1$ from below, indicating that the strength of radial NEC violation gradually decreases away from throat.
\item[(c)] The tangential EoS parameter $W_{\rm tan}$ remains greater
than $-1$ for parameter ranges displayed in panel~(b), consequently,
$\rho+P_{\rm tan}>0$, and tangential NEC is satisfied.
\item[(d)] The contrasting behavior of $W_{\rm rad}$ and $W_{\rm tan}$
reflects anisotropic character of effective matter distribution, while the exotic contribution required for traversability is therefore mainly
carried by radial pressure sector near throat.
\end{itemize}
Consequently, the EoS analysis supports existence of a traversable
Einasto dark matter supported WH in which radial NEC is violated in a
localized region around throat, while tangential NEC remains
satisfied for considered parameter choices, however this localized violation is consistent with requirement of exotic matter for maintaining  flare-out geometry without implying NEC violation in every spatial
direction.

\section{Total Amount of Exotic Matter via Volume Integral Quantifier in Wormhole Structure}
\label{ExoticMatter_VIQ}

\begin{figure}[h]
\centering
\includegraphics[height=5.4 in, width=6.4 in]{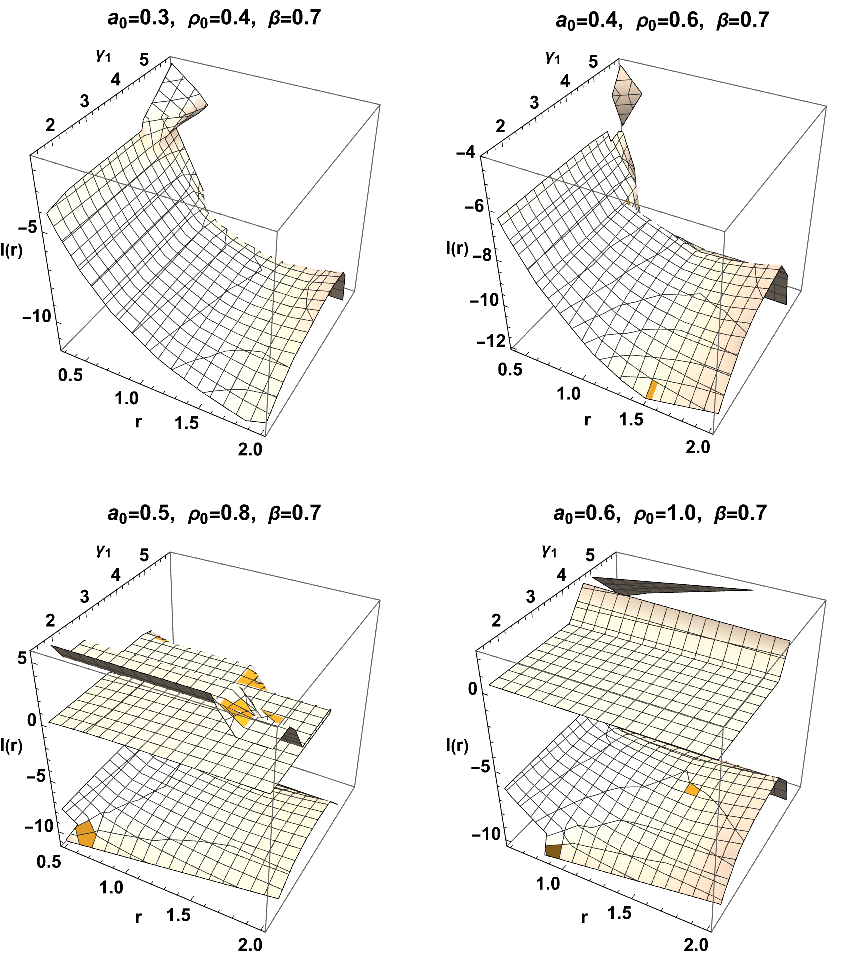}
\caption{Comparative three dimensional representation of volume
integral quantifier $I(r)$ as a function of radial coordinate $r$ and Einasto index $\gamma_{1}$ for four different combinations of WH
throat radius and central density: $(a_{0},\rho_{0})=(0.3,0.4)$, $(0.4,0.6)$, $(0.5,0.8)$, and $(0.6,1.0)$. Throughout the analysis, scale parameter and outer radial boundary are fixed at $\beta=0.7$ and $R_{1}=2$, respectively, however the four
panels illustrate how variations in $a_{0}$, $\rho_{0}$, and
$\gamma_{1}$ affect the accumulated radial NEC violating matter required
to sustain fuzzy WH configuration.}
\label{fig:viq_3d}
\end{figure}

A characteristic feature of traversable WH geometries in Einstein gravity is violation of NEC in a finite region surrounding throat, while in present model, radial NEC is violated whenever
$\rho+P_{\rm rad}<0,$ and this negative contribution characterizes exotic matter required to maintain flare-out geometry. Although the local behavior of $\rho+P_{\rm rad}$ identifies region in which NEC is violated, it does not directly quantify total amount of exotic matter distributed throughout WH spacetime. For this purpose, we employ volume integral quantifier, which provides an integrated measure of NEC violating matter content~\cite{visser2003traversable}. The volume integral quantifier is generally defined by integrating NEC combination over spatial volume,
\begin{equation}
I=\int \left(\rho+P_{\rm rad}\right)dV,
\end{equation}
where $dV$ denotes spatial volume element. For the static and spherically symmetric configuration considered here, the angular
integration can be performed explicitly and volume integral quantifier can be written as
\begin{equation}
I(r)=8\pi\int_{a_{0}}^{r}
\left(\rho+P_{\rm rad}\right)\tilde r^{\,2}\,d\tilde r,
\label{eq:VIQ}
\end{equation}
where $a_{0}$ is WH throat radius and $r>a_{0}$ denotes upper
radial limit of integration, while quantity $I(r)$ therefore measures the
cumulative contribution of the radial NEC combination between the throat
and a given radial position. For the present Einasto dark matter supported WH model, the explicit expression for $\rho+P_{\rm rad}$ has already been obtained in Eq.~\eqref{7a}, substituting this result into Eq.~(\ref{eq:VIQ}), we obtain
\begin{equation}
I(r,\gamma_{1},a_{0},\rho_{0},\beta)=8\pi\int_{a_{0}}^{r}\left(\rho+P_{\rm rad}\right)\tilde r^{\,2}\,d\tilde r .
\label{eq:VIQgeneral}
\end{equation}
The resulting integral contains nonlinear dependence of the
Einasto density profile through parameters $\gamma_{1}$, $\beta$, as well as $\rho_{0}$, together with throat radius $a_{0}$, since the resulting expression is not convenient for a compact analytical representation over parameter ranges of interest, volume integral quantifier is evaluated numerically. To examine influence of throat radius and the central density on exotic matter content, we consider the following four representative
parameter sets:
\begin{equation}
(a_{0},\rho_{0})=(0.3,0.4),\quad(0.4,0.6),\quad(0.5,0.8),\quad(0.6,1.0),
\end{equation}
while the remaining quantities are fixed as $\beta=0.7,\quad R_{1}=2$, for every pair $(a_{0},\rho_{0})$, the integral in
Eq.~(\ref{eq:VIQgeneral}) is evaluated over the interval
$r\in[a_{0},R_{1}],$
while the Einasto index $\gamma_{1}$ is varied in order to determine its
effect on accumulated exotic matter content. The numerical integration
is performed using an adaptive integration procedure, which is particularly suitable for the nonlinear form of the integrand generated by Einasto density distribution.
The resulting behavior of $I(r)$ as a function of radial coordinate
$r$ and the Einasto index $\gamma_{1}$ is displayed in Fig.~\ref{fig:viq_3d}. The four
panels correspond to four different combinations of $(a_{0},\rho_{0})$
specified above, while $\beta=0.7$ and $R_{1}=2$ are kept fixed throughout the analysis, however, graphical comparison allows us to determine how combined variation of throat radius, central density, radial distance, and Einasto index modifies total NEC violating matter content. Several important features can be extracted from Fig.~\ref{fig:viq_3d}:
\begin{itemize}
\item[(a)] Effect of the throat radius $a_{0}$: The magnitude of volume integral quantifier changes appreciably as the throat radius is
increased, while a larger throat generally requires the NEC violating matter to be distributed over a broader geometrical region, thereby modifying total exotic contribution necessary to sustain WH configuration.
\item[(b)] Effect of the central density $\rho_{0}$: The central density parameter directly controls strength and radial
localization of Einasto matter distribution, whereas for the parameter sets considered in Fig.~\ref{fig:viq_3d}, increasing $\rho_{0}$ modifies the magnitude as well as radial accumulation of $I(r)$ and can reduce the extent over which a strong negative NEC contribution persists.
\item[(c)] Dependence on the Einasto index $\gamma_{1}$: The parameter $\gamma_{1}$ controls the radial shape of the Einasto density
distribution and therefore has a direct influence on volume integral quantifier, whereas for smaller values of $\gamma_{1}$, NEC violating contribution can extend over a comparatively wider radial domain, allowing a larger negative contribution to accumulate. As $\gamma_{1}$ increases, the corresponding matter distribution becomes more localized and total exotic contribution is modified accordingly.
\item[(d)] Dependence on the radial coordinate $r$: Since $I(r)$ is an integrated quantity, its magnitude evolves as upper integration limit is moved outward from throat, however in those regions where
$\rho+P_{\rm rad}<0$, increasing $r$ incorporates additional
NEC violating matter into the integral. Once NEC violating contribution
becomes sufficiently weak away from throat, subsequent variation of
$I(r)$ is correspondingly reduced.
\end{itemize}
The comparative analysis of four parameter sets therefore demonstrates
that amount of exotic matter required to support the fuzzy WH is
sensitive to throat radius $a_{0}$, central density $\rho_{0}$,
and Einasto index $\gamma_{1}$, in particular, volume integral quantifier provides a useful global complement to local energy condition analysis by showing not only where radial NEC is violated but also how the corresponding negative contribution accumulates throughout WH interior. Furthermore, behavior of $I(r)$ supports interpretation that exotic matter content is predominantly associated with region close to WH throat, however this result is consistent with previously obtained
energy condition and EoS analyses, where radial NEC violation is mainly
confined to inner region whereas matter distribution becomes
progressively less exotic at larger radial distances. Consequently, volume integral quantifier analysis provides an additional quantitative diagnostic for assessing physical viability of Einasto dark matter supported fuzzy WH configuration.

\section{Conclusion}

In traditional WH theory, it is generally recognized that the existence of a physically admissible WH often demands a form of matter that leads to the violation of standard energy conditions, nevertheless, under EGR, certain configurations of the stress energy tensor can still comply with these conditions, opening the possibility for unconventional WH geometries. In this work, we concentrated on constructing spherically symmetric fuzzy WH models supported by anisotropic matter fields. Building on this foundation, we explore whether central galactic regions can be interpreted as fuzzy WH structures composed of dark matter governed by the Einasto density profile. 
\begin{description}
  \item[a)] We constructed and examined a new family of static, spherically symmetric fuzzy WH geometries within Einstein gravity, using the Einasto dark matter density profile as the matter source. The analytical form of the associated shape function was obtained and its characteristics analyzed through a series of plots in Fig.~\ref{flareout1} which confirm compliance with the fundamental WH criteria, namely the throat condition $B(a_0)=a_0$ and the flare-out requirement and other necessary requirements. 
  \item[b)] The radial and tangential EoS examined in our considered gravity, while the radial EoS parameter grows near the throat but exhibits a decreasing trend at larger radii, while tangential EoS parameter exhibits a distinct profile. The physical character of the matter threading the WH analyzed in detail, whereas the energy density remained positive at the throat and decreases sharply with increasing $r$, indicating that the exotic matter is confined to the innermost region. The radial pressure become negative near the throat, consistent with the requirements for WH geometries. The tangential pressure remained positive, while pressure anisotropy provided an outward-directed force contributing to the overall stability of our constructed fuzzy WH system.
  \item[c)] Energy condition demonstrated that both radial and tangential components become negative near the throat, indicating a localized violation of the NEC which is essential for traversability. 
  \item[d)] Visualization demonstration for stable behavior can be seen at throat radius $a_{0}=0.6$. The TOV analysis also confirms that hydrostatic and anisotropic forces counterbalance each anothers effects and ensuring equilibrium for appropriate parameter selections. Anisotropy parameter plot remains positive that counteracts gravitational collapse which supported stable behavior, while for a physically plausible configuration, average pressure must remain finite and positive. In our case it decreasing outward, as illustrated in Fig.~\ref{TOVAnAP}(c). 
  \item[e)] The complexity factor attains its maximum strength close to the WH throat as well as decreases progressively, approaching zero in the far field region which indicates that the non uniform distribution of matter and geometry, which gives rise to structural complexity, is mainly concentrated around the throat. The active gravitational mass, in contrast, takes slightly negative values near the throat due to the exotic matter contribution required for maintaining the WH geometry. However, it becomes positive at larger radial distances, reflecting the recovery of an attractive gravitational character away from the throat and the analysis is carried out for a fixed Einasto index $\gamma_{1}$.
  \item[f)] The WH geometry was further visualized using numerically constructed three dimensional embedding surfaces obtained by integrating the differential embedding equation in which each embedding diagram satisfies the throat condition $B(a_0)=a_0$, and Fig.~\ref{fig:combined3D} clearly demonstrates the presence of upper and lower sheets connected smoothly at the throat.
  \item[g)] A comparison across the three selected parameter sets reveals that larger values of  $\rho_{0}$ (increasing from 0.1 to 0.9) lead to a stronger concentration of matter near the throat, yielding a steeper neck and a quicker restoration of asymptotic flatness. Simultaneously, increasing \(\xi\) diminishes the magnitude of the nonlinear term through the $1/\xi^2$ factor, thereby reducing curvature away from the throat, so smaller $\rho_{0}$ which is index parameter and smaller $\xi$ result in a more extended curvature region, while on the other hand larger $\rho_{0}$ and larger $\xi$ restrict curvature to a narrow vicinity around $a_0$.
  \item[h)] The four three dimensional diagrams presented VIQ as a function of the radial coordinate $r$, while these plots emphasize the significance of this approach in assessing the physical plausibility of the proposed fuzzy WH configurations.
\end{description}
Throughout the data given in Figs. \ref{p1}-\ref{p4}, we can analyze the potential existence of stable and unstable configurations of the thin-shell with fuzzy black hole solutions to the WH geometry. The behavior of \(\zeta_0\) represents the positability of the stable zones around the critical points. Here, we analyze the effect of the dark fluid distribution on the stability of the shell using linearized radial perturbation. It can be seen that as parameter \(\xi\) increases, the shell becomes less stable (see Fig. \ref{p1}). In Fig. \ref{p2}, we analyze the effect of \(\gamma_1\) on the stability of the shell. It can be observed that the stable region increases before the critical points. For lower values of \(\beta\), we observe large stable regions (see Fig. \ref{p3}). As we increase \(\beta\), the region of stability decreases. For lower values of \(\rho_0\) (see Fig. \ref{p4}), we observe larger stable regions. The examination of escape trajectories shows that the monopole charge parameter, $\xi$, is fundamental to the analysis of the dynamics of test particles in WH geometries. Modified monopole charge affects the curvature of spacetime, and alters the potential contour around the throat. We show that a greater value of the angular momentum is associated with a stronger centrifugal barrier which results in a greater deflection and earlier escape. In contrast, a greater value of the particle energy leads to an earlier and quicker transition from quasi-bound state to unbound scattering trajectories. The combination of $\xi$, $L$, and $E$ determines the escape phenomena and the morphology of the orbits. These findings illustrate that the monopole charge affects the dynamics of particles, which can be linked to the nature of monopole-supported WH geometries. This analysis concludes that fuzzy WHs modeled with the Einasto dark matter density profile are consistent solutions within Einstein gravity, fulfilling critical features like throat formation, flare-out condition, and localized energy condition violations. Furthermore, the confinement of exotic effects near the throat and restoration of standard energy behavior at large radii suggest that the Einasto profile leads to a meaningful geometric structure for theoretical and astrophysical applications.

%\subsection*{Acknowledgment:}
%The authors extend their appreciation to the Deanship of Scientific Research at King Khalid University for funding this work through the Large Groups Project under grant number RGP.2/133/1445).

\vspace{0.5cm}

%\bibliography{bibm.yousaf}
%\bibliographystyle{apalike}
%\bibliographystyle{plain}
%\bibliographystyle{ieeetr}
%\bibliographystyle{plainnat}
%\bibliographystyle{abbrvnat}
%\bibliographystyle{unsrtnat}

\end{document}